\documentclass[prb,twocolumn,aps,amsmath,amssymb]{revtex4}
\usepackage[T1]{fontenc}
\usepackage[utf8]{inputenc}
\usepackage{graphicx}
\usepackage{amsmath}
\usepackage{amssymb}
\usepackage{dcolumn}
\usepackage{float}
\usepackage{bm}
\usepackage{amsfonts,times}
 \usepackage[breaklinks=true,colorlinks,citecolor=blue,linkcolor=blue,urlcolor=blue]{hyperref}

\DeclareMathAlphabet{\bi}{OML}{cmm}{b}{it}

\def\be{\begin{equation}}
\def\ee{\end{equation}}
\def\bearr{\begin{eqnarray}}
\def\eearr{\end{eqnarray}}
\def\la{\langle}
\def\ra{\rangle}

\begin{document}
\title{Probing Topological signatures in an optically driven
$\bm \alpha$-$\bm{T_3}$ Lattice}

\author{Lakpa Tamang}
\author{Tutul Biswas}
\email{tbiswas@nbu.ac.in}
\normalsize
\affiliation
{Department of Physics, University of North Bengal, Raja Rammohunpur-734013, India}
\date{\today}

\begin{abstract}
The $\alpha$-$T_3$ lattice, an interpolation model between the honeycomb lattice of
graphene($\alpha=0$) and the dice lattice($\alpha=1$), undergoes a topological phase transition across $\alpha=1/\sqrt{2}$ when exposed to a circularly polarized off-resonant light. We study Berry phase mediated bulk magnetic and anomalous thermoelectric responses in order to capture the topological signatures of a driven $\alpha$-$T_3$ lattice. It is revealed that both the Berry curvature and the orbital magnetic moment associated with the flat band change their respective signs across $\alpha=1/\sqrt{2}$. The off-resonant light distorts the flat band near the Dirac points when $0<\alpha<1$ which eventually introduces two distinct well separated forbidden gaps of equal width in the quasienergy spectrum. The orbital magnetization varies linearly with the chemical potential, in the forbidden gaps. The slopes of the linear regions in the orbital magnetization are closely related to the respective Chern numbers on either side of $\alpha=1/\sqrt{2}$. We find that the slope for $\alpha>1/\sqrt{2}$ is approximately two times of that for  $\alpha<1/\sqrt{2}$ which essentially indicates a topological phase transition across $\alpha=1/\sqrt{2}$. However, the anomalous Nernst coefficient vanishes when the chemical potential is tuned in the forbidden gaps. The anomalous Hall conductivity in the forbidden gap(s) approaches different quantized values on either side of 
$\alpha=1/\sqrt{2}$. All these topological signatures can be observed experimentally. As a consequence of the broken particle-hole symmetry for $0<\alpha<1$, we find valley-contrasting features in the thermoelectric coefficients as well as in the orbital magnetization which further open up the possibility to use the underlying driven system in the valley caloritronic applications. In addition, we obtain some analytical results in the case of the irradiated dice lattice$(\alpha=1)$. Interestingly, we find that the undistorted flat band of the driven dice lattice contributes nothing to the Berry curvature. However, the flat band gives rise to a finite orbital magnetic moment which is the sum of the contributions due to the conduction and the valence band.   
 
\end{abstract}

\maketitle
\section{Introduction}
Based on the Berry phase effect[\onlinecite{BP_Pardgm}], a new paradigm in condensed matter physics has been developed over last few decades. The Berry phase[\onlinecite{M_Berry}] is a global phase acquired by a Bloch electron while its adiabatic expedition along a closed path in momentum space. It is expressed as a line integral of the Berry connection: ${\bm A}(\bm k)=i\la u({\bm k})\vert \bm \nabla_{\bm k}\vert u(\bm k)\ra$ over a closed contour in momentum space, where $\vert u(\bm k)\ra$ is the periodic part of Bloch wave function. The Berry curvature: ${\bm \Omega}(\bm k)=\bm\nabla_{\bm k}\times \bm A(\bm k)$ plays an analogous role of magnetic field in the momentum space. In the modern theory of condensed matter physics, the Berry phase effect has been regarded as a unifying concept in order to understand a wide variety of intriguing phenomena such as the topological origin of the quantum Hall effect[\onlinecite{topo_qhe}], quantized adiabatic pumping[\onlinecite{qnt_pump}], the anomalous Hall effect[\onlinecite{ahe1, ahe2, ahe3, ahe4}], electric polarization[\onlinecite{El_polarz}], orbital Magnetization[\onlinecite{OM_Semi, OM_Wann1, OM_Wann2, OM_Pert}], spin transport[\onlinecite{spin_trans1, spin_trans2}], the valley Hall effect[\onlinecite{VHE1, VHE2}] etc. The topological thermoelectric phenomena[\onlinecite{Ano_Therm}] mediated by the Berry phase has gained considerable interest over the years. Several systems, such as spin-chiral ferromagnetic Kagome lattice[\onlinecite{Nernst_Kagome}], graphene[\onlinecite{Nernst_Grphn}], 
spin-orbit coupled electron/hole doped two-dimensional(2D) semiconductors[\onlinecite{Nernst_Rashba}], transition metal dichalcogenides[\onlinecite{Nernst_MoS2, Nernst_WSe2}], topological insulator[\onlinecite{Nernst_topo}], Weyl semimetals[\onlinecite{Nernst_Weyl1, Nernst_Weyl2, Nernst_Weyl3}] etc exhibit topological signatures in their thermoelectric response, especially in the Nernst coefficients. It is revealed that a gap in the low energy spectrum despite of its origin is an essential ingredient to observe the topological Nernst effect, especially in the 2D systems. The Nernst effect provides an experimental platform to investigate a wide variety of phenomena such as the detection of vortex-like excitations[\onlinecite{vortex_Ex}] and charge density waves[\onlinecite{Cdw}] in cuprate superconductors, anomalous thermoelectric response of graphene[\onlinecite{NG1,NG2,NG3}], large Berry curvatures in Dirac semimetals[\onlinecite{BC_Probe}] etc. In addition, the Nernst effect has potential technological application in the field of spin caloritronics[\onlinecite{spin_Calr}].

The possibility of tuning the low-energy properties of electronic systems dynamically with the help of an external time periodic radiation has opened a new pathway in condensed matter physics nowadays. The notion ``Floquet systems" is used synonymously with the periodically driven systems because the non-equilibrium dynamics of the underlying systems are well understood within the framework of the Floquet theory[\onlinecite{Floquet}]. The radiation-matter interaction in such systems has enormous consequence in the sense that it can induce non-trivial topology in the band structure. The study on light-induced Hall effect in graphene by Oka et al[\onlinecite{Oka_gr}] triggered a number of subsequent investigations aiming to probe topological phase transition[\onlinecite{Tanaka1, Lindner1, Lindner2, Oka2, Tanaka2, Ezawa, Moessner, Platero, Usaj, Jin, Saha, Xing1}],
Floquet spin states[\onlinecite{Schliemann1}], pseudospin effects[\onlinecite{Schliemann2}], dynamical localization[\onlinecite{dyn_boson, dyn_Gr}], chiral interfacial modes[\onlinecite{Firoz1}], spin-Hall resonance without a magnetic field[\onlinecite{Firoz2}] etc in several irradiated systems. More specifically, a circularly polarized light of frequency lying in the off-resonant regime (frequency is high enough compared to any other relevant frequency scale of the system) is able to renormalize the band structure through a second-order virtual photon emission-absorption process. Finally, one gets an effective static Hamiltonian with a gap term whose nature entirely depends on the properties of the external radiation e.g. intensity, frequency and polarization state. This tunable gap is the central reason behind all photo-engineered topological phases as mentioned in the above references. For example, the semimetallic graphene can be transformed into a Chern insulator when illuminated by off-resonant light of appropriate frequency[\onlinecite{Oka_gr}]. To check how these externally induced topological flavors essentially control the transport properties, a plethora of investigations have been performed on several irradiated systems[\onlinecite{Oka2,Floq_Trans2,Floq_Trans3,Floq_Trans4,tahir1,tahir2,Floq_Trans5,Floq_Trans6,Floq_Trans7,Floq_Trans8,
Floq_Trans9, Floq_Trans10,Debashree,Floq_Trans11,Floq_Trans12}] in recent years.

In this work, we extract some topological flavors of an irradiated $\alpha$-$T_3$ lattice by analyzing the Berry phase mediated magnetic and thermoelectric transport properties. The 
$\alpha$-$T_3$ lattice is identical to a honeycomb lattice or a dice lattice[\onlinecite{dice1,dice2,dice3}] when the variable
$\alpha\in[0,1]$ becomes  $\alpha=0$ or 
$\alpha=1$, respectively. Several schemes[\onlinecite{Prop1,Prop2,Prop3}] have been proposed for a possible experimental realization of the $\alpha$-$T_3$ lattice. It is a fast evolving topic of current research as it hosts quasiparticles having pseudospin larger than $S=1/2$. The Berry phase associated with the adiabatic evolution of the eigenstates depends on $\alpha$ explicitly. This fact is itself manifested in a number  of fascinating phenomena including orbital magnetic susceptibility[\onlinecite{Mag_Suscp}], Hall quantization[\onlinecite{T3_Hall1, T3_Hall2}],
Klein tunneling[\onlinecite{Klein1, Klein2}], Weiss oscillations[\onlinecite{Weiss}], zitterbewegung[\onlinecite{ZB}], plasmons[\onlinecite{Plasmon1,Plasmon2, Plasmon3, Plasmon4}], peculiar magneto-optical effects[\onlinecite{Mag_Opt1, Mag_Opt2, Mag_Opt3, Mag_Opt4}], Ruderman-Kittel-Kasuya-Yosida interaction[\onlinecite{RKKY1, RKKY2}], minimal conductivity[\onlinecite{Min_Con}], unconventional topology[\onlinecite{Ghosh_Topo}], spin-Hall phase transition[\onlinecite{spin_Hall_T3}] etc. When exposed to a time periodic external radiation, the $\alpha$-$T_3$ lattice exhibits rich physical phenomena which are reported in a series of recent works[\onlinecite{Bashab1, Bashab2, Iurov_Floq, Mojarro, TB_Floq}]. In particular, the irradiated system supports a topological phase transition 
across $\alpha=1/\sqrt{2}$\,[\onlinecite{Bashab2}]. Given this background, it is therefore tempting to investigate the transport properties of an irradiated $\alpha$-$T_3$ lattice aiming to extract some topological signatures therein. Very recently, Niu and Wang[\onlinecite{Trans_aT3}] numerically studied the behaviour of the valley polarized current in a circularly polarized light irradiated $\alpha$-$T_3$ lattice using the nonequilibrium Green's function formalism. However, the topological aspects of the valley current went largely unaddressed. Here, we study the explicit behaviors of the Berry curvature, the orbital magnetic moment, the orbital magnetization, and anomalous thermoelectric coefficients such as the Nernst coefficient, the anomalous Hall conductivity and the thermal Hall conductivity. It is revealed that all the aforesaid quantities exhibit distinct topological signatures. For example, both the Berry curvature and the orbital magnetic moment associated with the flat band changes their respective signs across $\alpha=1/\sqrt{2}$. The linear variation of the orbital magnetization and the vanishing anomalous Nernst coefficient when the chemical potential encounters the forbidden gaps are indeed direct topological signatures. Particularly, the slopes of the linear regions in the orbital magnetization are directly related to the Chern numbers on either side of $\alpha=1/\sqrt{2}$. The anomalous Hall conductivity also exhibits topological features when the chemical potential is varied in the forbidden gap(s). We also find valley contrasting features in the observables for $0<\alpha<1$ as a direct consequence of the broken particle-hole symmetry. In addition, we obtain some analytical results in the case of an irradiated dice lattice. Our results are experimentally detectable and have potential applications in the field of valley caloritronics. 

The rest of the paper is organized as follows. In section II, we outline the basic formalism and discuss general features of the quasienergy band structures, the Berry curvature, the orbital magnetic moment, and the orbital magnetization associated with the irradiated $\alpha$-$T_3$ lattice. Various aspects of the topological thermoelectric transport are discussed in section III. This article ends with a summary given in section IV.

\section{Formalism} 
We consider the $\alpha$-$T_3$ lattice being exposed to a circularly polarized light as shown in 
Fig.\,\ref{fig:Fig_Lattice}(a). We start our discussion by reviewing the essential features of  its quasienergy dispersions. We, then, systematically discuss various characteristics of the  Berry curvature, the orbital magnetic moment and the orbital magnetization.

\subsection{Quasienergy}
\begin{figure}[h!]
\centering
 \includegraphics[width=8.5cm, height=4.0cm]{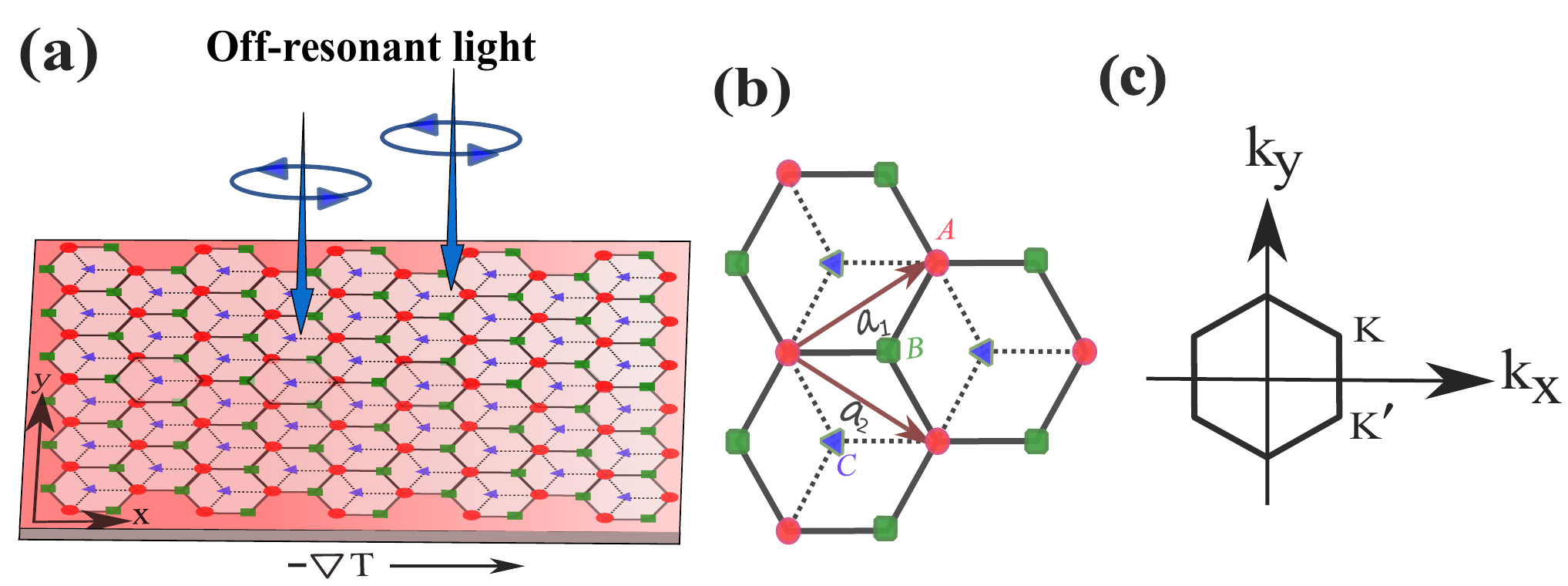}
 \caption{(\textcolor{blue}{Color Online}) (a) The bulk portion of the irradiated $\alpha$-$T_3$ lattice is schematically shown here. A circularly polarized light is incident on the lattice ( located in the $x$-$y$ plane) normally. The system is subjected to a weak temperature gradient $-\bm \nabla T$ along the $x$-direction. (b) The geometry of the $\alpha$-$T_3$ lattice is sketched here. (c) The first Brillouin zone of the 
$\alpha$-$T_3$ lattice. Here, $K$ and $K^\prime$ are the valleys.}
\label{fig:Fig_Lattice}
\end{figure}

As shown in Fig.\,\ref{fig:Fig_Lattice}(b), a unit cell of the $\alpha$-$T_3$ lattice includes three inequivalent lattice sites, namely, $A$, $B$, and $C$. The site $C$, at the center of each hexagon in the honeycomb structure formed by $A$ and $B$ sites, is connected to three $A$ sites only. The lattice translational vectors are given by
$\bm a_1=(3/2, \sqrt{3}/2)a$ and $\bm a_2=(3/2, -\sqrt{3}/2)a$, where $a$ is 
the inter-site distance. An electron at site $B$\,$(C)$ can hop to the site $A$ with an energy cost $\gamma$\,$(\alpha\gamma)$, where 
$\alpha\in[0,1]$. Therefore, $\alpha=0$\,$(\alpha=1)$ corresponds to the case of graphene\,(dice lattice). Within the nearest neighbor tight-binding scenario, the static $\alpha$-$T_3$ model supports a zero energy flat band on which the conduction and the valence band touch each other at six Dirac points. Only two of them are independent which represent, furthermore, the valleys $K$ and $K^\prime$ as shown in Fig. \ref{fig:Fig_Lattice}(c).  
The low energy excitations near a Dirac point in a particular valley are governed by the Hamiltonian
\begin{eqnarray}
H_0^{\tau}(\bm{k})=
\begin{pmatrix}
0&f^{\tau}_{\bm{k}} \cos\phi&0\\
f^{\tau^\ast}_{\bm{k}} \cos\phi&0&f^{\tau}_{\bm{k}}\sin\phi\\
0&f^{\tau^\ast}_{\bm{k}}\sin\phi&0
\end{pmatrix}, 
\end{eqnarray}
where
$f^{\tau}_{\bm{k}}=\hbar v_F(\tau k_x - ik_y)$
with $v_F=3\gamma a/(2\hbar)$ being the Fermi
velocity, $\tau=\pm 1$ is the valley index and $\phi=\tan^{-1}(\alpha)$.

\begin{figure}[h!]
\centering
 \includegraphics[width=10cm, height=6.5cm]{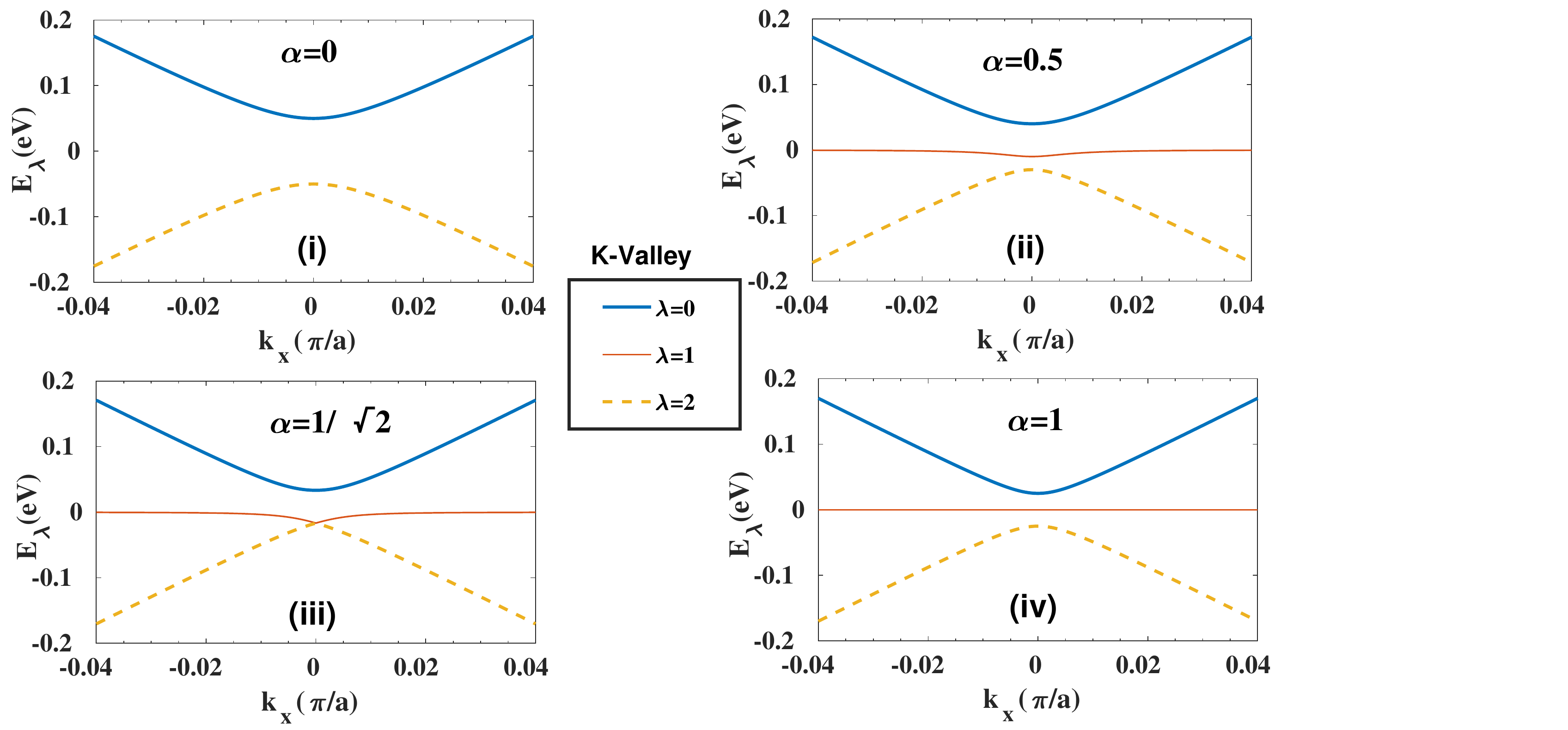}
 \caption{(\textcolor{blue}{Color Online}) Quasienergy dispersion is shown for the $K$-valley corresponding to different values of $\alpha$. Here, we consider a right circularly polarized light ($l=+1$) with $\Delta=50$ meV. It is understood that an external time periodic radiation lifts the degeneracy at ${\bm k}=0$ for all values of $\alpha$ except $\alpha=1/\sqrt{2}$. At this particular value of $\alpha$, the valence band touches the distorted flat band.}
\label{fig:Fig_Ek}
\end{figure}

We consider a circularly polarized light of frequency 
$\omega$ described by the vector potential ${\bm{A}(t)}=A_0(l\cos\omega{t},\sin\omega{t})$ is incident normally on the $\alpha$-$T_3$ lattice. Here, $A_0$ is the amplitude and $l$ represents the polarization of the light. Now, the minimal coupling between the external radiation and the system modifies the Hamiltonian further as  
\begin{eqnarray}\label{mod_Ham}
H^{\tau}(\bm{k},t)=
\begin{pmatrix}
0&f^{\tau}_{\bm{k}}(t) \cos\phi&0\\
f^{\tau^\ast}_{\bm{k}}(t) \cos\phi&0&f^{\tau}_{\bm{k}}(t)\sin\phi\\
0&f^{\tau^\ast}_{\bm{k}}(t)\sin\phi&0
\end{pmatrix},
\end{eqnarray}
where $f^{\tau}_{\bm k}(t)$ is obtained from $f^{\tau}_{\bm k}$ via Pierl's substitution: $\bm{k}\rightarrow\bm{k}+e\bm{A}(t)/\hbar$.
The modified Hamiltonian $H^\tau({\bm k},t)$ has the same periodicity $T=2\pi/\omega$ as ${\bm A}(t)$. The Floquet theory perhaps is the appropriate tool to deal with such time-periodic problem. We consider the so-called off-resonant regime where the frequency of the radiation is much larger than the band width and/or other energy scales of the system under consideration. In this case, one can obtain the following  
time-independent effective Hamiltonian[\onlinecite{Magnus_Fl}] 
\begin{eqnarray}\label{Eff1}
H^{\tau}_{\rm eff}(\bm{k})=H^{\tau}_0(\bm{k})+\frac{1}{\hbar\omega}
\Big[H^{\tau}_-(\bm{k}),H^{\tau}_+(\bm{k})\Big]
+\mathcal{O}(1/{\omega^2}),
\end{eqnarray}
where $H^{\tau}_{\pm}(\bm{k})=(1/T){\int_{0}^{T}e^{\mp {i}{\omega}t}{H^{\tau}(\bm{k},t)}}\,dt$. The second term in Eq.\,(\ref{Eff1}) is responsible for the virtual photon emission-absorption process which effectively alters the static band structure. Considering terms up to $\mathcal{O}(1/\omega)$, it is straightforward to find the effective Hamiltonian as
\begin{eqnarray}
H^{\tau}_{\rm eff}(\bm{k})=
\begin{pmatrix}
\Delta^{\tau} \cos^2\phi & f^{\tau}_{\bm{k}}\cos\phi & 0 \\
f^{\tau^\ast}_{\bm{k}}\cos\phi & -\Delta^{\tau} \cos2\phi & f^{\tau}_{\bm{k}}\sin\phi\\
0 & f^{\tau^\ast}_{\bm{k}}\sin\phi & -\Delta^{\tau} \sin^2\phi
\end{pmatrix}
\end{eqnarray}
where,
$\Delta^{\tau}$=$\tau l\Delta$ with $\Delta$=${e^2A_{0}^2v_F^2/\hbar\omega}$ is the Haldane-type mass term induced by the off-resonant light. It breaks the time reversal symmetry and has opposite signs in the two valleys. Diagonalization of $H_{\rm eff}^\tau(\bm k)$  provides the quasienergies as
\begin{eqnarray}\label{QsEn}
E^{\tau}_\lambda(\bm{k})=2\sqrt{\frac{-p}{3}} \cos\Bigg[\frac{1}{3}\cos^{-1}\Bigg(\frac{3q}{2p}\sqrt{\frac{-3}{p}}\Bigg)-\frac{2\pi \lambda}{3}\Bigg],
\end{eqnarray}
 where
\begin{eqnarray}
&& p=-\Big[{\vert f_{\bm{k}}\vert}^2+\Delta^2\Big(\cos^2 2\phi+\frac{\sin^2 2\phi}{4}\Big)\Big]\nonumber\\
&&q=-\tau l\frac{\Delta^3\sin^2 2\phi \cos2\phi}{4}\nonumber.
\end{eqnarray}
Here, $\lambda$=0, 1, and 2 represent quasienergies corresponding to the conduction, flat, and valence bands, respectively.
Eq.\,(\ref{QsEn}) infers that the application of an off-resonant radiation makes the effective quasienergy dispersion $\alpha$ dependent.
The nature of the quasienergy dispersion at the $K$-valley for different values of $\alpha$ are depicted in Fig.\,\ref{fig:Fig_Ek}. We consider the system is being exposed to a right circularly polarized radiation ($l=+1$) of which the amplitude and the frequency are chosen in a such way that $\Delta=50$ meV.
The effect of the circularly polarized radiation on the band structure of the system is mainly two-fold. Firstly, it breaks the time reversal symmetry. Therefore, the three-fold degeneracy at the Dirac point ($\bm k=0$) is lifted by opening a gap. However, the scenario corresponding to
$\alpha=1/\sqrt{2}$ is different. Here, the valence band touches the flat band. More specifically, a topological phase transition occurs at 
$\alpha=1/\sqrt{2}$\,[\onlinecite{Bashab2}]. Secondly, the radiation distorts the flat band in the vicinity of the Dirac point for an intermediate $\alpha$ ($0<\alpha<1$). As a result, the particle-hole symmetry is broken. 
However, the aforesaid symmetry is still preserved for graphene($\alpha=0$) as well as for the dice lattice($\alpha=1$). For instance, the quasienergy spectrum for an illuminated dice lattice becomes : $E_\pm^{\rm d}(\bm k)=\pm(\varepsilon_k^2+{\tilde{\Delta}}^2)^{1/2}$, $E_0^{\rm d}=0$, where
$\varepsilon_k=\hbar v_Fk$ and $\tilde{\Delta}=\Delta/2$. In other words, the ``flatness'' of the flat band of a dice lattice is protected against the application of high-frequency radiation. 
We mention, here, that the particle-hole transformation is associated with the replacement of an electron with wavevector $\bm k$ by a hole with wavevector $\bm k$.

\begin{figure}[h!]
\centering
 \includegraphics[width=10cm, height=6.5cm]{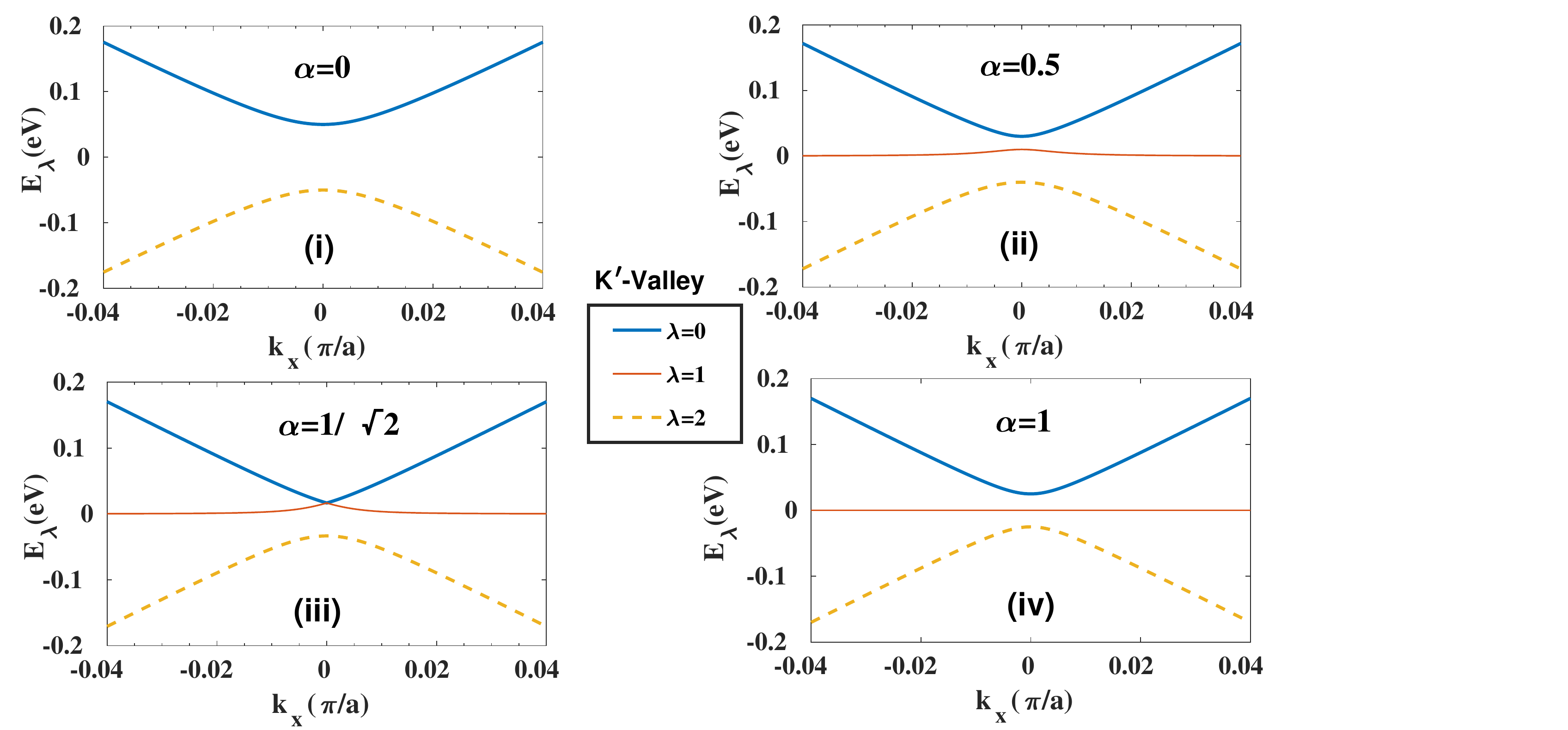}
 \caption{(\textcolor{blue}{Color Online}) We repeat Fig.\,\ref{fig:Fig_Ek} for the $K^\prime$-valley. Here, it is revealed that the conduction band touches the distorted flat band at $\alpha=1/\sqrt{2}$.}
\label{fig:Fig_Ekp}
\end{figure}

In Fig.\,\ref{fig:Fig_Ekp}, we have shown the quasienergy dispersion for $K^\prime$-valley for the same parameters. In this case, the conduction band touches the distorted flat band at $\alpha=1/\sqrt{2}$. The feature corresponding to the $K^\prime$-valley can also be realized in the $K$-valley by reversing the polarization of radiation ($l=-1$). 

\subsection{Berry Curvature}
\begin{figure}[h!]
\centering
\includegraphics[width=10cm, height=6.5cm]{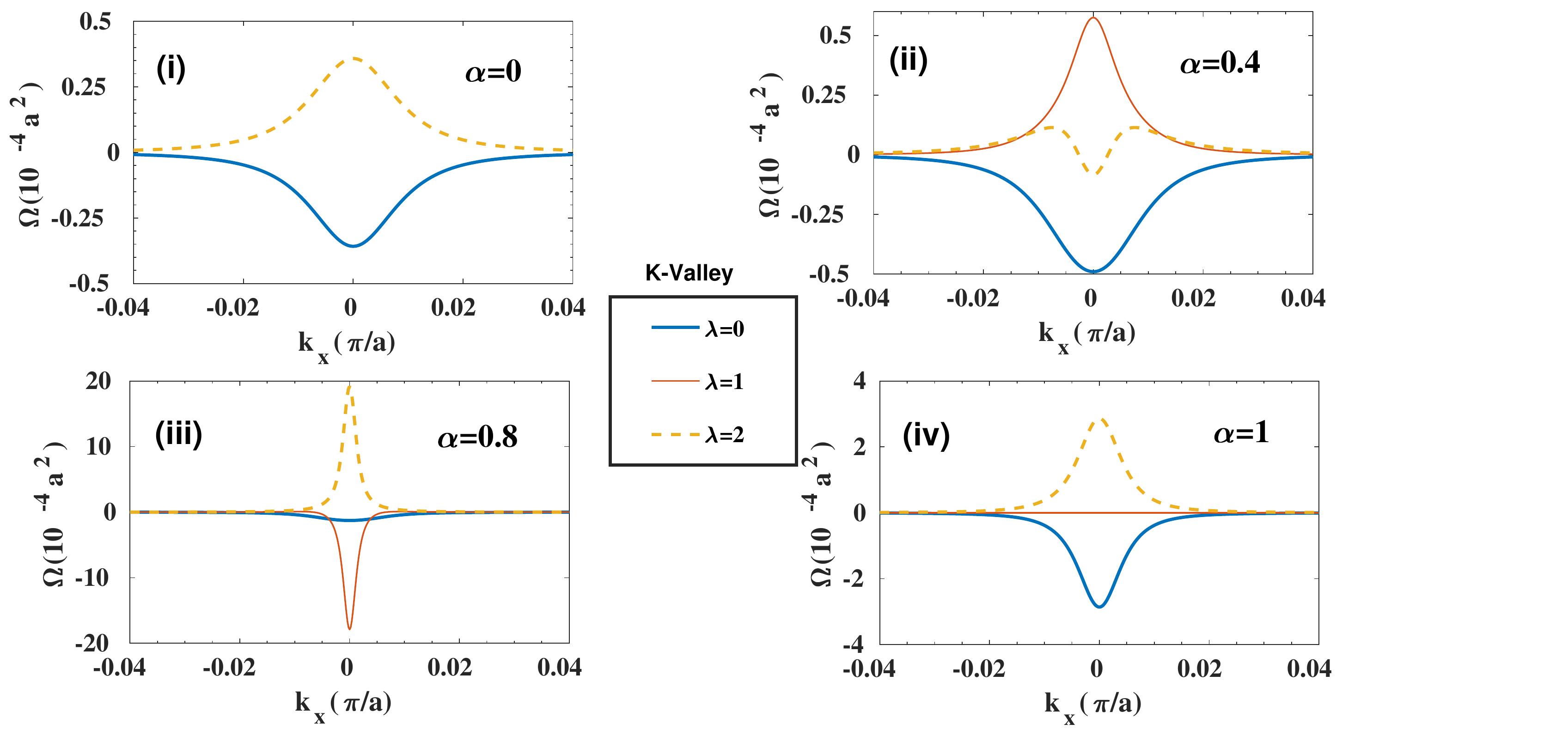}
\caption{(\textcolor{blue}{Color Online}) The distributions of the Berry curvature $\Omega(\bm k)$ near the 
$K$-valley for different values of $\alpha$, namely, (i) $\alpha=0$, (ii) $\alpha=0.4$, (iii) $\alpha=0.8$, and (iv) $\alpha=1$ are shown. Here, we consider a right circularly polarized light ($l=+1$) with $\Delta=50$ meV to irradiate the $\alpha$-$T_3$ lattice.}
\label{fig:Fig_BC_K}
\end{figure}

The Berry curvature corresponding to a particular Bloch band can be viewed as a kind of ``residual'' interaction[\onlinecite{BP_Pardgm}] of other nearby bands as the dynamics of the system is locked to a single energy band within the description of the quantum adiabatic theorem. It is well known that the Berry curvature for a two-dimensional system is always directed along the transverse direction. Its gauge-invariant form corresponding to a 
Bloch band characterized by the indices $\lambda$ and $\tau$ is given by
\begin{eqnarray}\label{Berry_C}
{\Omega}_{\lambda}^{\tau}(\bm{k})=-2\hbar^2\, {\rm Im}\sum\limits_{\substack{\lambda^\prime\neq\lambda}}
{\frac{{\langle u_\lambda^\tau\vert v_x\vert{u_{\lambda^\prime}^{\tau}}\rangle\langle u_{\lambda^\prime}^{\tau}\vert v_y \vert{u_\lambda^\tau}\rangle}}{\left[E_\lambda^\tau(\bm{k})-E_{\lambda^\prime}^{\tau}(\bm{k})\right]^2}},
\end{eqnarray}
where $u_{\lambda}^\tau\equiv \vert u_{\lambda}^\tau(\bm k)\ra=\sqrt{S}e^{-i\bm k\cdot \bm r}\vert\Psi_\lambda^\tau(\bm k)\ra$ with $S$ as the area of the sample and 
$v_i=\hbar^{-1}\nabla_{\bm k_i}H_{\rm eff}^\tau$ is the effective velocity operator along a particular direction $i=x,y$. It is evident from
Eq.\,(\ref{Berry_C}) that the Berry curvature becomes singular when there is a degeneracy in the energy spectrum. Under the spatial 
inversion($\mathcal{I}$), the particle-hole($\mathcal{P}$) and the time-reversal($\mathcal{T}$) operations, 
the Berry curvature behaves in the following way:   
$\mathcal{I}^{-1} \Omega_\lambda(\bm k)\mathcal{I}=\Omega_\lambda(-\bm k)$, 
$\mathcal{P}^{-1}\Omega_{\bar{\lambda}}(\bm k)\mathcal{P}=-\Omega_\lambda(-\bm k)$, and
$\mathcal{T}^{-1}\Omega_\lambda(\bm k)\mathcal{T}=-\Omega_\lambda(-\bm k)$, respectively. Here, the index
$\bar{\lambda}$ corresponds to the band with energy $-E_\lambda$. A non-vanishing $\Omega({\bm{k}})$, therefore, demands the breaking of at least one of above mentioned discrete symmetries. For the irradiated $\alpha$-$T_3$  lattice, the behavior of the Berry curvature around the $K$-valley is shown in Fig.\,\ref{fig:Fig_BC_K}. We consider $l=+1$ and $\Delta=50$ meV. The Berry curvature for an individual band becomes non-vanishing as a consequence of broken time-reversal symmetry for all values of $\alpha$. In all cases, 
$\Omega({\bm k})$ is mostly concentrated near the valley extremum. The Berry curvature for the conduction band is negative for all values of 
$\alpha$. It is hard to comment on the topological features from the behavior of $\Omega(\bm k)$ for the conduction band. The Berry curvature for the valence band, however, exhibit non-monotonic behavior. 
For $\alpha=0$, it is positive and peaked at $\bm k=0$. A ``cusp"-like structure with a negative peak value appears when $\alpha$ becomes $0.4$. For $\alpha=0.6$(not shown here) the Berry curvature becomes negative. The Berry curvature is strongly enhanced and becomes positive when $\alpha=0.8$. It is still positive if one increases $\alpha$ further to $\alpha=1$. Here, $\Omega(\bm k)$ changes the sign around $\alpha=1/\sqrt{2}$ which might be considered as a topological signature. For $0<\alpha<1$, the Berry curvature corresponding to the flat band is non-vanishing as a consequence of the particle-hole symmetry breaking. Interestingly, 
$\Omega(\bm k)$ for the flat band exhibits a sign change across $\alpha=1/\sqrt{2}$. For the dice lattice($\alpha=1$), the contribution of the flat band in $\Omega(\bm k)$ vanishes as the external radiation is unable to break the particle-hole symmetry. It is possible to obtain following analytical expression of
$\Omega(\bm k)$ for $\alpha=1$ and $l=+1$ as
\begin{eqnarray}\label{Berry_C_Dice}
\bm{\Omega}_\lambda^\tau(\bm{k})=(\lambda-1)\frac{\hbar^2v_F^2\tilde{\Delta}}{(\varepsilon_k^2+\tilde{\Delta}^2)^{3/2}}.
\end{eqnarray}

\begin{figure}[h!]
\centering
\includegraphics[width=8.5cm, height=4cm]{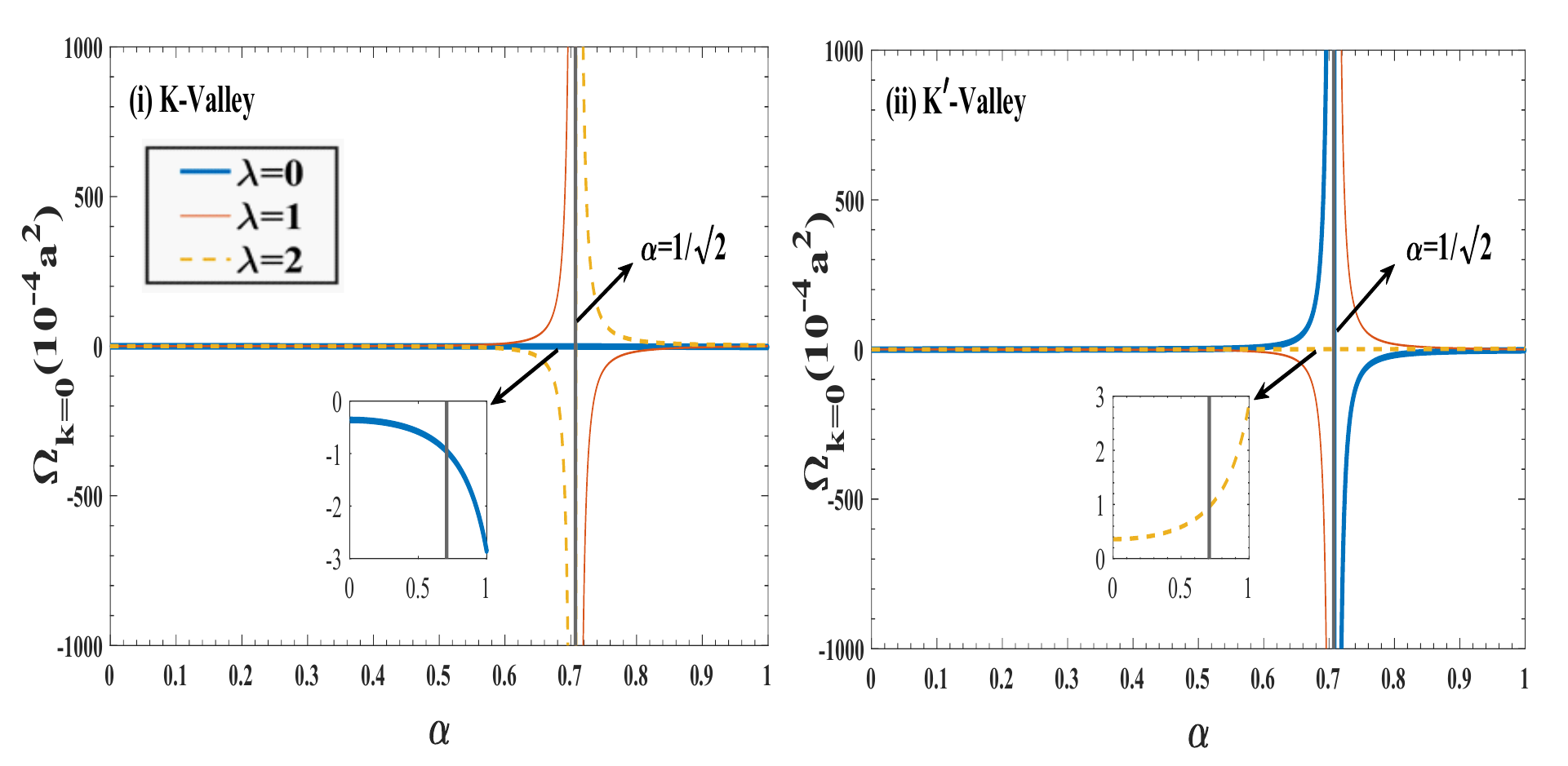}
\caption{(\textcolor{blue}{Color Online}) The peak value of $\Omega(\bm k)$ at $\bm k=0$ is plotted as a function of $\alpha$ for 
(i) the $K$-valley and (ii) the $K^\prime$-valley. Here, we consider $l=+1$ and $\Delta=50$ meV. In the 
$K$\,($K^\prime$)-valley, $\Omega(\bm k=0)$ for the flat\,(conduction) band and the valence\,(flat) band change signs discontinuously across 
$\alpha=1/\sqrt{2}$ whereas $\Omega(\bm k=0)$ for the conduction\,(valence) band decreases\,(increases) monotonically as depicted in the inset. The sign change of the Berry curvature across $\alpha=1/\sqrt{2}$ might be considered as a topological signature.}
\label{fig:Fig_PeakBC}
\end{figure}

It is also clear that, at a given $\bm k$, the total Berry curvature, i.e. the sum of the individual contributions from different bands, vanishes. This is usually known as local conservation of the Berry curvature. 
In Fig.\,\ref{fig:Fig_PeakBC}(i)(ii), the peak value of $\Omega(\bm k)$ at $\bm k=0$ is depicted over the entire range of $\alpha$ for the $K$($K^\prime$) valley. For the $K$-valley, the Berry curvatures corresponding to the flat band and the valence band diverge at 
$\alpha=1/\sqrt{2}$ while the Berry curvature for the conduction band is finite(as shown in the inset). This divergence is a direct consequence of the fact that the valence band and the flat band touch each other at $\bm k=0$ when $\alpha=1/\sqrt{2}$ as shown in Fig.\,\ref{fig:Fig_Ek}(iii). As mentioned earlier, 
$\Omega({\bm k=0})$ for the flat\,(valence) band changes sign from 
$+\,(-)$ to $-\,(+)$ as $\alpha$ is varied across $\alpha=1/\sqrt{2}$ in order to exhibit topological signature. For the $K^\prime$-valley, $\Omega({\bm k=0})$ is finite for the valence band while that corresponding to the conduction band and the flat band encounter divergence at $\alpha=1/\sqrt{2}$ because the conduction band and the flat band become degenerate at $\bm k=0$. The Berry curvatures for the conduction band and the flat band changes their respective signs across $\alpha=1/\sqrt{2}$ as depicted in 
Fig.\,\ref{fig:Fig_PeakBC}(ii).  

\subsection{Orbital Magnetic moment}
Another interesting quantity associated with the Bloch band of a given system is the orbital magnetic moment(OMM). The self rotation of an electronic wave packet about the center of mass gives rise to the OMM. The OMM exhibits analogous behavior as the electron spin. In principle, it can be treated as a physical observable because various informations about it can be extracted by studying the magnetic circular dichroism spectrum[\onlinecite{MCD1, MCD2}]. Generally, it is expressed as
\begin{eqnarray}
\bm{m}_\lambda^\tau(\bm{k})=-\frac{ie}{2\hbar}\langle\nabla_{\bm{k}} u_\lambda^\tau\vert\times[H_{\rm eff}^\tau(\bm{k})-E_\lambda^\tau(\bm{k})]\vert\nabla_{\bm{k}}u_\lambda^\tau\rangle.
\end{eqnarray} 

The $z$-component of the OMM, however, can be obtained as
\begin{eqnarray}
{m}_{\lambda}^{\tau}(\bm{k})=-\hbar e\, {\rm Im} \sum\limits_{\substack{\lambda^\prime
\\(\lambda\neq\lambda^\prime)}}
{\frac{{\langle u_\lambda^\tau\vert v_x\vert{u_{\lambda^\prime}^{\tau}}\rangle\langle u_{\lambda^\prime}^{\tau}\vert v_y \vert{u_\lambda^\tau}\rangle}}{\left[E_\lambda^\tau(\bm{k})-E_{\lambda^\prime}^{\tau}(\bm{k})\right]}}.
\end{eqnarray}

\begin{figure}[h!]
\centering
\includegraphics[width=10cm, height=6.5cm]{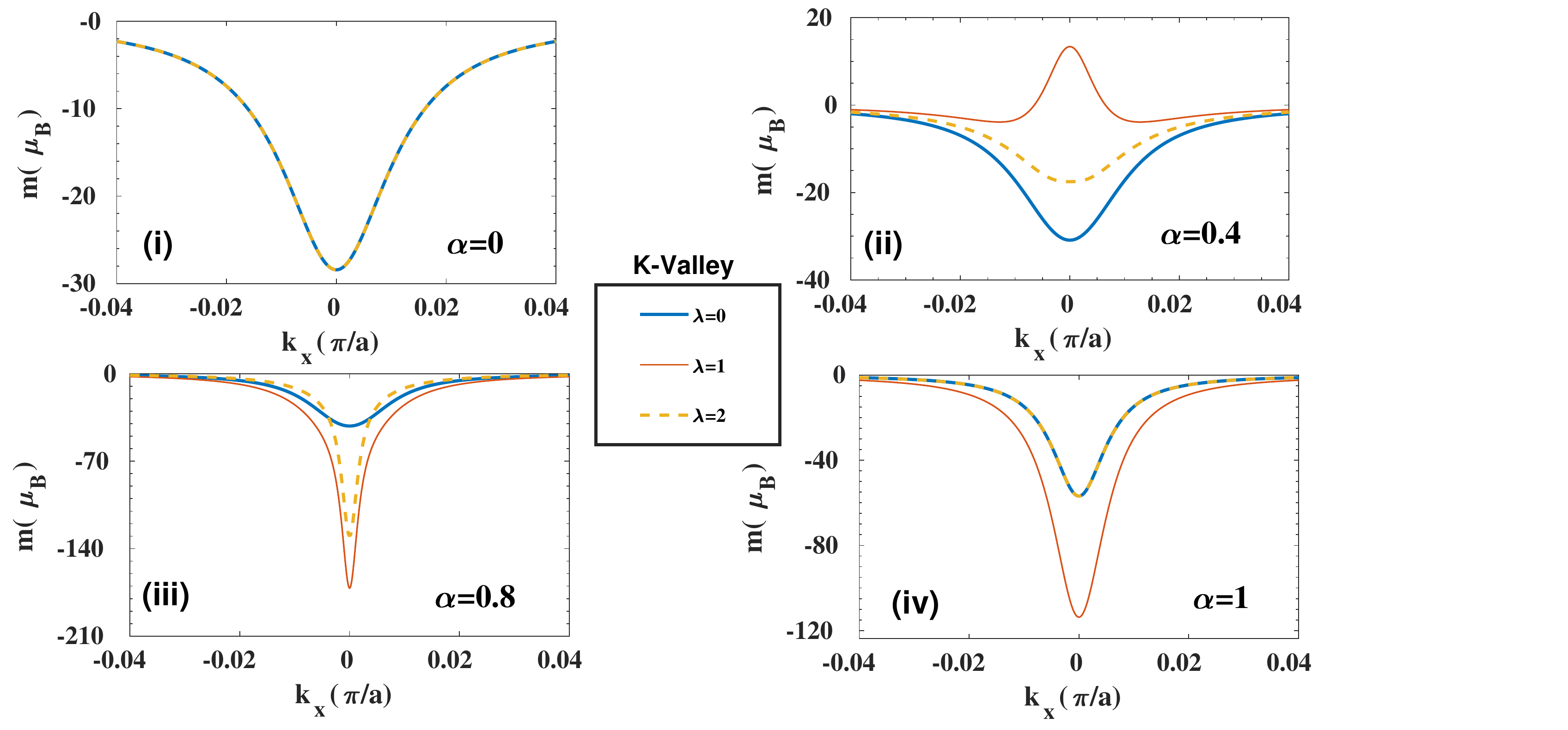}
\caption{(\textcolor{blue}{Color Online}) The distribution of the OMM around the $K$-valley for different values of $\alpha$. Here, we choose $l=+1$ and $\Delta=50$ meV. For $\alpha=1$, the flat band's contribution to the OMM does not vanish. Moreover, it is exactly equal to the sum of the individual contributions of the conduction and valence bands.}
\label{fig:Fig_MagMt}
\end{figure}

The distribution of the OMM is shown in Fig.\,\ref{fig:Fig_MagMt} for the $K$-valley. The symmetry properties of the OMM are as follows:    
$\mathcal{I}^{-1} m_\lambda(\bm k)\mathcal{I}=m_\lambda(-\bm k)$, 
$\mathcal{P}^{-1}m_{\bar{\lambda}}(\bm k)\mathcal{P}=m_\lambda(-\bm k)$, and
$\mathcal{T}^{-1}m_\lambda(\bm k)\mathcal{T}=-m_\lambda(-\bm k)$.
The OMM is largely concentrated around the valley extremum (i.e. $\bm k\approx0$) like the Berry curvature. However, it exhibits some distinct features which are absent in the distribution of the Berry curvature. The broken time-reversal symmetry gives rise to a non-zero OMM. For $\alpha=0$ and $\alpha=1$, the OMMs associated with the conduction band and the valence band are negative and they coincide with each other owing to the particle-hole symmetry. For the 
dice lattice ($\alpha=1$), we find that the OMM associated with the flat band is non-vanishing unlike the Berry curvature and it is exactly the sum of the individual contributions due to the conduction band and the valence band. This is indeed an interesting result. We find the OMM analytically for $\alpha=1$ as 
\begin{eqnarray}
{m}_\lambda^\tau (\bm{k})=-\frac{\hbar ev_F^2\tilde{\Delta}}{2(\varepsilon_k^2+\tilde{\Delta}^2)}
\Big(\delta_{\lambda 0}+2\delta_{\lambda 1}+\delta_{\lambda 2}\Big).
\end{eqnarray}

\begin{figure}[h!]
\centering
 \includegraphics[width=8.5cm, height=4cm]{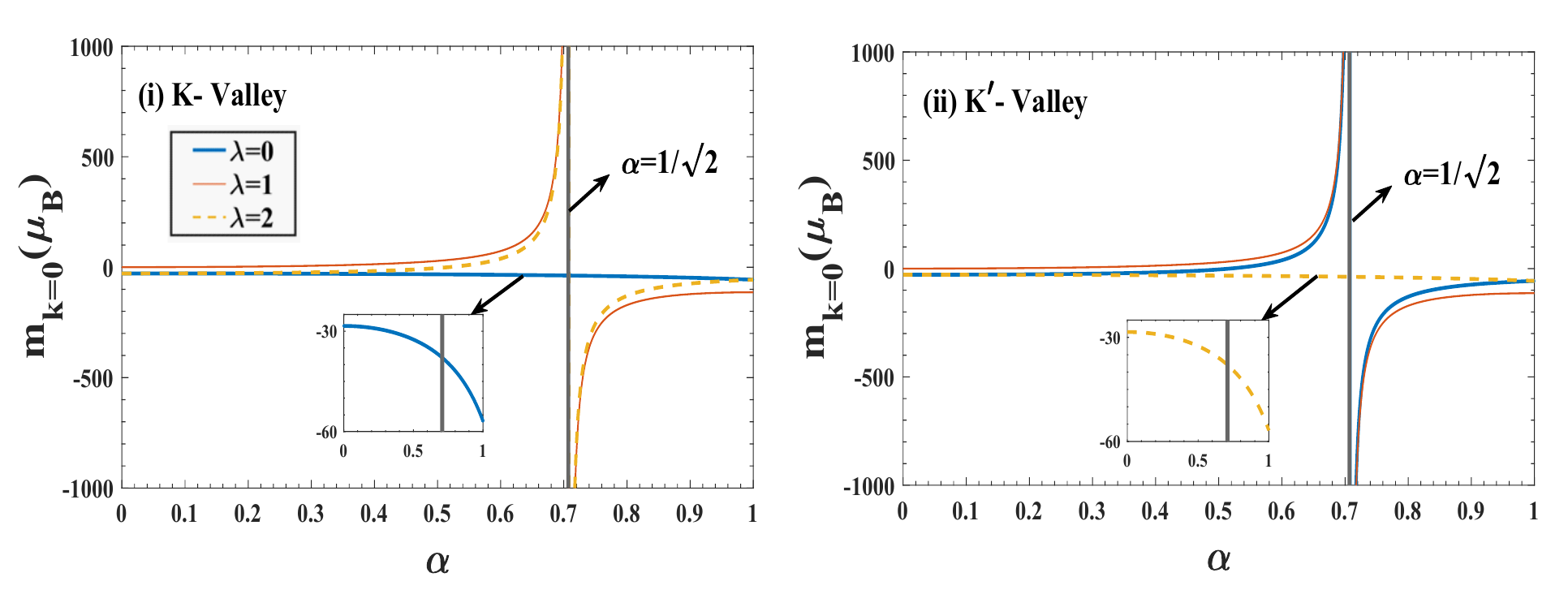}
 \caption{(\textcolor{blue}{Color Online}) The peak value of $ m(\bm k)$ at $\bm k=0$ is plotted as a function of $\alpha$. 
(i) In the $K$-valley, the OMMs due to the flat band and the valence band discontinuously change their signs across $\alpha=1/\sqrt{2}$ whereas that due to the conduction band decreases monotonically with
$\alpha$ (as shown in the inset). (ii) The role played by the conduction band in (i) is replaced by the valence band in case of the $K^\prime$-valley.}
\label{fig:Fig_MagMt_Peak}
\end{figure}

For an intermediate value of $\alpha$ i.e. $0<\alpha<1$, the breaking of the inversion, the particle-hole, and the time-reversal symmetries result in different values of the OMM associated with individual bands. To explore the topological features of the OMM, we show the variation of $m_\lambda^\tau({\bm k}=0)$ with
$\alpha$ for both valleys in Fig.\,\ref{fig:Fig_MagMt_Peak}. At $K$-valley, the OMMs corresponding to the flat band and the valence band change their respective signs across $\alpha=1/\sqrt{2}$ while that due to the conduction band decreases monotonically with $\alpha$. For the $K^\prime$-valley, the role of the valence band is replaced by the conduction band. 

\subsection{Orbital Magnetization}
The orbital magnetization is an interesting bulk property of crystalline materials in which the time reversal symmetry is broken. In its modern understanding based on either the semiclassical wave packet dynamics of 
Bloch electrons[\onlinecite{OM_Semi}] or the 
Wannier function approach[\onlinecite{OM_Wann1, OM_Wann2}] or the perturbation theory[\onlinecite{OM_Pert}], it is revealed that 
the orbital magnetization is comprised of two contributions due to the OMM and the Berry curvature, separately.
The orbital magnetization is simply the derivative of free energy with respect to the external magnetic field. The free energy of the system in presence of a weak magnetic field $\bm{B}$ is given by 
\begin{eqnarray}
F^\tau=-\frac{1}{\beta} \sum_{\lambda,{\bm{k}}}
{\rm ln}\Big[(1+e^{-{\beta(\varepsilon_{\lambda}^\tau(\bm k)-\mu)}}\Big].
\end{eqnarray} 
Here, $\beta=1/(k_BT)$, $k_B$ is the Boltzmann constant, $T$ is the temperature and $\mu$ is the chemical potential. Note that the band energy
$E_{\lambda}^\tau$ is modified to $\varepsilon_{\lambda}^\tau(\bm k)=E_\lambda^\tau
-\bm{m}_\lambda^\tau\cdot\bm{B}$  as a result of the coupling between the OMM and the magnetic field.

In presence of the Berry curvature, the summation over $\bm k$ can be converted into an integral as[\onlinecite{OM_Semi}]
\begin{eqnarray}\label{dos_b}
\sum_{\bm k}\longrightarrow \frac{1}{(2\pi)^2}\int \Bigg(1+\frac{e{\bm B}\cdot{\bm \Omega}_\lambda^\tau(\bm k)}{\hbar}\Bigg)\,d^2k.
\end{eqnarray}
The Berry curvature essentially modifies the phase-space density of states(the second term in 
Eq.\,(\ref{dos_b})) as a consequence of the violation of Liouville's theorem in connection with the conservation of the phase-space volume[\onlinecite{OM_Semi}]. 

The orbital magnetization for a particular valley is given by 
$M_{\rm orb}^\tau=-(\partial F^\tau/\partial {B})_{\mu, T}$ which can be further obtained as 
$ M_{\rm orb}^\tau=M_{\rm avg}^\tau+M_{\rm com}^\tau$, where
\begin{eqnarray}
&& M_{\rm avg}^\tau=\frac{1}{(2\pi)^2}\sum_\lambda\int m_\lambda^\tau(\bm{k})f_\lambda^\tau(\bm{k})d^2k, \label{M_av}\\
&& M_{\rm com}^\tau=\frac{e}{2\pi\beta{h}}\sum_\lambda\int\Omega_\lambda^\tau(\bm{k}) \ln\big[1+e^{\beta(\mu-E_\lambda^\tau(\bm{k}))}\big]d^2k.\nonumber\label{M_cm}\\
\end{eqnarray}
Here, $f_\lambda^\tau(\bm{k})=\{1+\exp[\beta(E_\lambda^\tau(\bm k)-\mu)]\}^{-1}$ is the Fermi-Dirac distribution function and the integrations in 
Eq.\,(\ref{M_av}) and Eq.\,(\ref{M_cm}) are over the states below the chemical potential $\mu$. Note that  $M_{\rm avg}$ is just the thermodynamic average of the OMM and $M_{\rm com}$ is the Berry phase mediated extra term associated with the center of mass motion of the wave packet.

It is possible to find analytical expressions of the orbital magnetization in the case of the irradiated dice lattice ($\alpha=1$) at very low temperatures. In the limit $T\rightarrow0$, when $\mu$ stays in the 
conduction($+$)/valence($-$) band, we obtain $M_{\rm orb}$ as, 
\begin{eqnarray}\label{Morb_P}
M_{\rm orb}^\pm=\mp\frac{e\mu}{h}\Bigg(1-\frac{\tilde{\Delta}}{\sqrt{\mu^2+\tilde{\Delta}^2}}\Bigg)+\frac{e\tilde{\Delta}}{4h}\ln{\Big |\frac{\mu^2+\tilde{\Delta}^2}{\tilde{\Delta}^2}}\Big |.
\end{eqnarray}

\begin{figure}[h!]
\centering
 \includegraphics[width=9cm, height=5cm]{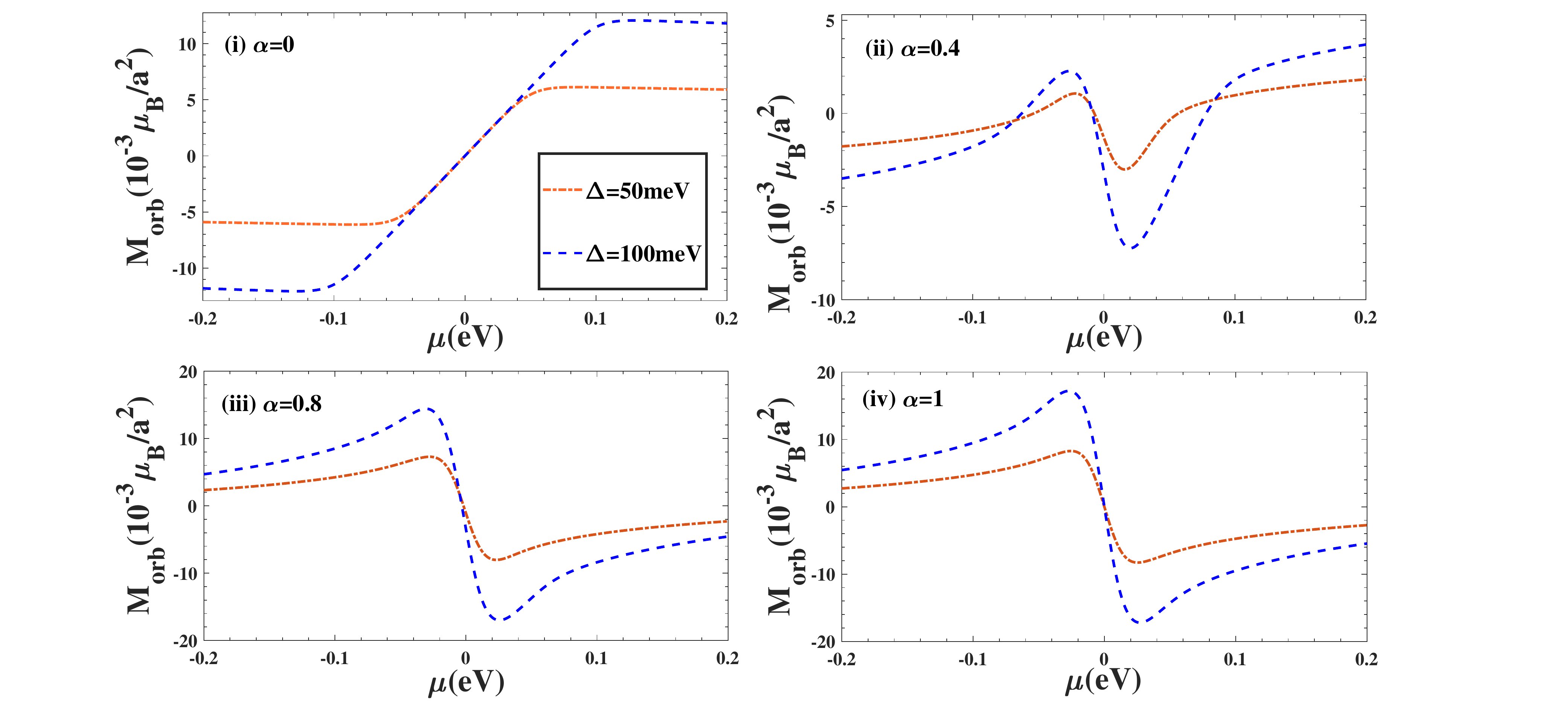}
 \caption{(\textcolor{blue}{Color Online}) The orbital magnetization at the $K$-valley as a function of the chemical potential for various values of $\alpha$. We consider $l=+1$ and $T=100$ K. $M_{\rm orb}$ behaves in anti-symmetric manner with $\mu$ when $\alpha=0$ and 1. For $0<\alpha<1$, the broken particle-hole symmetry causes a deviation from this behavior.}
\label{fig:Fig_Orb_Mag}
\end{figure}

For an intermediate $\alpha$ i.e. $0<\alpha<1$, however, the $\bm k$-integrations in Eq.\,(\ref{M_av}) and Eq.\,(\ref{M_cm}) are evaluated numerically to understand the behavior of $M_{\rm orb}$. 
In Fig.\,\ref{fig:Fig_Orb_Mag}, the variation of $M_{\rm orb}$ with the chemical potential $\mu$ are shown for various values of $\alpha$, namely, 
$\alpha=0, 0.4, 0.8,$ and $1$ at $T=100$ K. We consider two values of the light induced energy gap, namely, $\Delta=50$ meV and 
$\Delta=100$ meV. A higher
$\Delta$ enhances the magnitude of $M_{\rm orb}$ significantly for all values of $\alpha$. 
For $\alpha=0$ [Fig.\,\ref{fig:Fig_Orb_Mag}(i)] and 
$\alpha=1$ [Fig.\,\ref{fig:Fig_Orb_Mag}(iv)], $M_{\rm orb}$ changes anti-symmetrically with $\mu$ despite of completely different $\mu$-dependencies. For example, 
$M_{\rm orb}$ switches from a negative (positive) value to a positive (negative) value around $\mu=0$ when $\alpha=0\,(1)$. 
These anti-symmetric natures are absent in the cases of $\alpha=0.4$ [Fig.\,\ref{fig:Fig_Orb_Mag}(ii)] and $\alpha=0.8$ [Fig.\,\ref{fig:Fig_Orb_Mag}(iii)] as a consequence of broken particle-hole symmetry.

\begin{figure}[h!]
\centering
 \includegraphics[width=9.5cm, height=5.5cm]{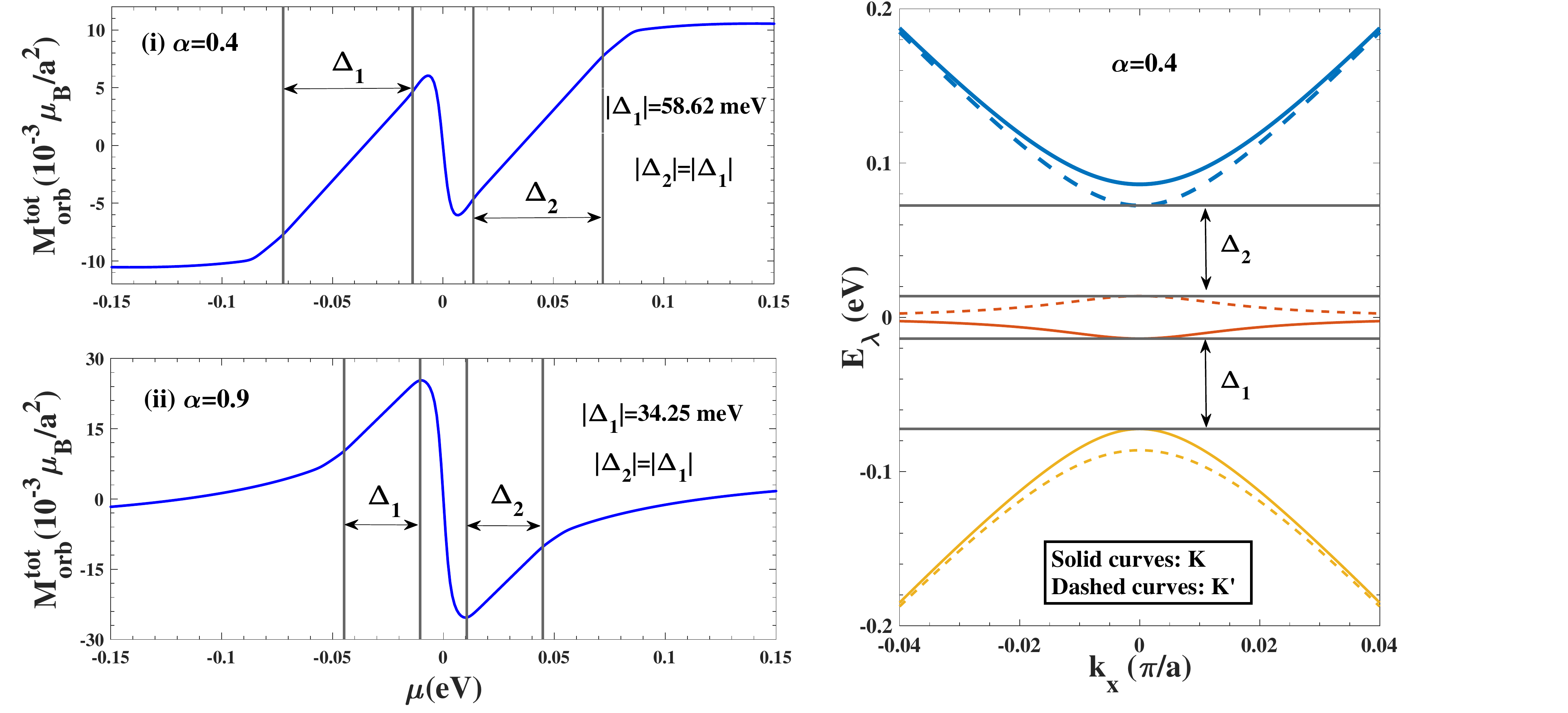}
 \caption{(\textcolor{blue}{Color Online}) (Left Panel) The $\mu$-dependence of the total orbital magnetization i.e. sum of the contributions from both valleys for (i) $\alpha=0.4$ and (ii) $\alpha=0.9$. (Right Panel) Quasienergy dispersion for both valleys when $\alpha=0.4$.  Here, we consider $l=+1$, $T=15$\,K and
$\Delta=100$ meV.}
\label{fig:Fig_Orb_Mag_T15_tot}
\end{figure}

To extract the topological flavors, we show the $\mu$-dependence of the total orbital magnetization which is the sum of contributions from both valleys in the left panel of Fig.\,\ref{fig:Fig_Orb_Mag_T15_tot}.
We choose lower temperature and higher $\Delta$, namely, $T=15$ K and $\Delta=100$ meV in order to visualize the topological signatures in the orbital magnetization.
The left panel of Fig.\,\ref{fig:Fig_Orb_Mag_T15_tot} reveals that $M_{\rm orb}^{\rm tot}$ varies linearly with $\mu$ in two well separated ``windows" $\Delta_1$ and $\Delta_2$ of equal width. Interestingly,
$\Delta_1$ is the energy gap between the flat band and the valence band at the $K$-valley whereas 
$\Delta_2$ is the band gap between the conduction band and the flat band at the $K^\prime$-valley as depicted in the right panel of Fig.\,\ref{fig:Fig_Orb_Mag_T15_tot}. Note that the widths of the ``windows" strongly depend on $\alpha$. We find $\Delta_1=58.62$ meV when $\alpha=0.4$ and $\Delta_1=34.25$ meV for 
$\alpha=0.9$. The linear portions of $M_{\rm orb}^{\rm tot}$ are determined by the Berry phase mediated term 
$M_{\rm com}^{\rm tot}$. We have checked that (not shown here)
$M_{\rm avg}^{\rm tot}$ exhibits plateaus of different heights when $\mu$ encounters the band gaps. 
The linear variation of $M_{\rm orb}^{\rm tot}$ with $\mu$ in the forbidden gap(s) is indeed a topological signature which can be understood from the following relation[\onlinecite{OM_Wann2}]:
\begin{eqnarray}\label{M_gap}
\frac{dM_{\rm orb}^{\rm tot}}{d\mu}=\frac{e}{h}\sum_\lambda^{\rm occ}C_\lambda,
\end{eqnarray}
where the summation is over the occupied bands and $C_\lambda$ is the Chern number corresponding to the quasienergy band index $\lambda$. As mentioned earlier, an irradiated
$\alpha$-$T_3$ lattice undergoes a topological phase transition across
$\alpha=1/\sqrt{2}$\,[\onlinecite{Bashab2}] which is characterized by a change in the Chern number from 
$(C_0, C_1, C_2)=(-1, 0, 1)$ to $(C_0, C_1, C_2)=(-2, 0, 2)$. Therefore, the cases corresponding to $\alpha=0.4$ and $\alpha=0.9$ are topologically distinct. We have estimated the slopes of the linear portions of $M_{\rm orb}$ in the forbidden gaps $\Delta_1$ and 
$\Delta_2$ from Fig.\,\ref{fig:Fig_Orb_Mag_T15_tot} as well as from Eq.\,(\ref{M_gap}). A comparison between the obtained results is presented in the Table I.
\begin{table}[ht]
\centering
\begin{tabular}{|c|c|c|c|c|}\hline

\multicolumn{1}{|c|}{Value of} &
\multicolumn{2}{c|}{~~~~Slope in $\Delta_1$~~~~} &
\multicolumn{2}{c|}{~~~~Slope in $\Delta_2$~~~~} \\ \cline{2-5}

\multicolumn{1}{|c|}{$\alpha$} &
\multicolumn{1}{|c|}{~~Fig.\,\ref{fig:Fig_Orb_Mag_T15_tot}~~} &
\multicolumn{1}{|c|}{~~Eq.\,\ref{M_gap}~~} &
\multicolumn{1}{|c|}{~~Fig.\,\ref{fig:Fig_Orb_Mag_T15_tot}~~} &
\multicolumn{1}{|c|}{~~Eq.\,\ref{M_gap}~~} \\ \hline
\ \ \ 0.4 & 0.83 & 1 & 0.83 & 1 \\
\hline
\ \ \ 0.9 & 1.83 & 2 & 1.83 & 2 \\
\hline

\end{tabular}
\caption{The slopes of the linear portions of $M_{\rm orb}^{\rm tot}$ in units of $e/h$ in the ``windows": $\Delta_1$ and $\Delta_2$ calculated from Fig.\,\ref{fig:Fig_Orb_Mag_T15_tot} and Eq.\,(\ref{M_gap}) are tabulated here. Note that for a given $\alpha$, the slopes of $M_{\rm orb}^{\rm tot}$ in $\Delta_1$ and $\Delta_2$ are same which can be attributed to the vanishing Chern number of the flat band.}
\end{table}
It is noteworthy that the slopes of the linear regions in $\Delta_1$ and $\Delta_2$ are same for both values of $\alpha$. This can be explained as follows. For instance, for 
$\alpha=0.4$, when $\mu$ is varied in $\Delta_1$, only the valence band is occupied for which the Chern number is $1$. Then, Eq.\,(\ref{M_gap}) confirms the linear variation of $M_{\rm orb}^{\rm tot}$ with $\mu$ in the energy gap 
$\Delta_1$.  On the other hand, when $\mu$ is tuned in the gap 
$\Delta_2$, both the flat and the valence bands are occupied. As the Chern number corresponding to the flat band is $0$, it is easy to conclude that 
$M_{\rm orb}^{\rm tot}$ should vary linearly with $\mu$ with the same slope as that in $\Delta_1$. Similar argument will also hold in the case of
$\alpha=0.9$. 
The ratio of the slopes for $\alpha=0.9$ and $\alpha=0.4$ in either band gap obtained from Fig.\,\ref{fig:Fig_Orb_Mag_T15_tot} is approximately $2.20$ which is in close agreement with that obtained from Eq.\,(\ref{M_gap}). 
The peculiar behavior of $M_{\rm orb}^{\rm tot}$ when 
$\mu$ is tuned in the region between the ``windows" $\Delta_1$ and
$\Delta_2$ is entirely attributed to the distorted flat band.

\begin{figure}[h!]
\centering
 \includegraphics[width=9.5cm, height=6.5cm]{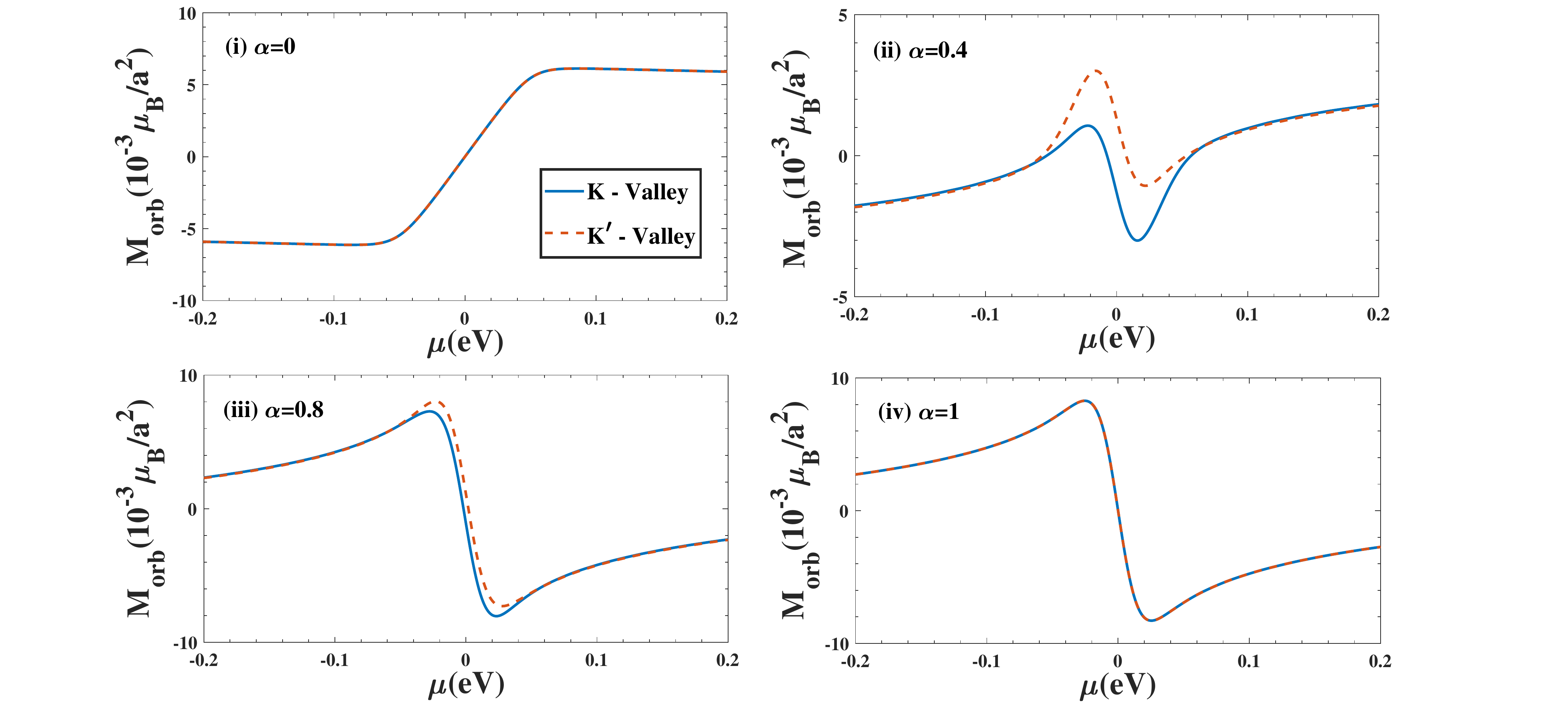}
 \caption{(\textcolor{blue}{Color Online}) Orbital Magnetization as function of $\mu$ for both valleys. Here, we consider $\Delta=50$ meV, $T=100$ K, and $l=+1$. }
\label{fig:Fig_Morb_KKp}
\end{figure}
In Fig.\,\ref{fig:Fig_Morb_KKp}, we show the dependence of $M_{\rm orb}$
on $\mu$ for both valleys considering $\Delta=50$ meV, $T=100$ K, and $l=+1$. For $\alpha=0$ and $\alpha=1$, we find that $M_{\rm orb}$ at the $K$-valley coincides with that at the $K^\prime$-valley as a result of the particle-hole symmetry. However, for $0<\alpha<1$, the breaking of both particle-hole and time reversal symmetries lead to a valley contrasting $M_{\rm orb}$ as shown in
Fig.\,\ref{fig:Fig_Morb_KKp} (ii) and (iii).

\section{Anomalous thermoelectric coefficients}   
In this section, we intend to study the Berry-phase mediated Nernst-Ettinghausen and Righi-Leduc effects in an irradiated $\alpha$-$T_3$ lattice. We focus on the behavior of relevant thermoelectric coefficients, mainly the Nernst and the thermal conductivity tensors $\overleftrightarrow{\alpha}$ and $\overleftrightarrow{\kappa}$, respectively. The conventional Nernst effect is associated with the generation of a transverse voltage in the presence of a temperature gradient and an external magnetic field. However, it is possible to detect the Nernst signal in the absence of a magnetic field. This is usually known as the anomalous Nernst effect(ANE). Here, a non-trivial Berry curvature plays a role of an effective magnetic field in the reciprocal space so that the charge carrier gets a transverse anomalous velocity. One can manipulate the finite spread of a wave packet representing a charge carrier to develop a semiclassical theory of anomalous thermoelectric transport phenomena. It is demonstrated that a Berry-phase correction term in the orbital magnetization plays an important role in 
the ANE[\onlinecite{Ano_Therm}]. For a particular valley, the expressions for the anomalous Nernst coefficient(ANC), and the thermal Hall conductivity(THC), are, respectively, given by[\onlinecite{Ano_Therm, Diss_Nernst}]
\begin{eqnarray}\label{ANC}
\alpha_{xy}^\tau &=&-\frac{k_Be}{h}\sum_{\lambda} \int \frac{d^2k}{(2\pi)^2}\, \bm\Omega_\lambda^\tau(\bm{k})\Big\{\beta(E_\lambda^\tau-\mu)f_\lambda^\tau(\bm{k})\nonumber\\
&+& \, \ln\big[1-f_\lambda^\tau(\bm{k})\big]\Big\},
\end{eqnarray}
and 
\begin{eqnarray}\label{THC}
\kappa_{xy}^\tau &=& \frac{k_B^2T}{h}\sum_{\lambda}\int \frac{d^2k}{(2\pi)^2}\,\bm\Omega_\lambda^\tau(\bm{k})\Bigg\{\frac{\pi^2}{3}+\beta^2(E_\lambda^\tau-\mu)^2 f_\lambda^\tau(\bm{k})\nonumber\\
 &-& 2{\rm Li}_2\big[1-f_\lambda^\tau(\bm{k})\big]
-\Big[\ln(1+e^{-\beta(E_\lambda^\tau-\mu)}\Big]^2 \Bigg\}.
\end{eqnarray}
Here, ${\rm Li}_2(z)$ is the polylogarithmic function. The quantity within the curly bracket in Eq.\,(\ref{ANC}) can be identified as the entropy density $S_\lambda^\tau(\bm k)=-f_\lambda^\tau(\bm{k})\, \ln[f_\lambda^\tau(\bm{k})]-[1-f_\lambda^\tau(\bm{k})]\,\ln[1-f_\lambda^\tau(\bm{k})]$. An entropy generation around the Fermi surface and the Berry curvature both can control the behavior of
$\alpha_{xy}$. Therefore, $\alpha_{xy}$ becomes very much sensitive to any changes in the Fermi surface properties such as the Fermi energy, temperature etc. However, the Berry curvature alone determines the anomalous Hall conductivity(AHC) as given by[\onlinecite{An_HE1, An_HE2}]
\begin{eqnarray}\label{AHC}
\sigma_{xy}^\tau=\frac{e^2}{\hbar}\sum_\lambda \int \frac{d^2k}{(2\pi)^2}\, \Omega_{\lambda}^\tau(\bm k)f_\lambda^\tau(\bm k).
\end{eqnarray}

In the $T\rightarrow0$ limit, Eq.\,(\ref{ANC}) and Eq.\,(\ref{THC}) reduce to the following Mott relation and Widemann-Franz law, respectively,
\begin{equation}
\alpha_{xy}^\tau=-\frac{\pi^2k_B^2T}{3e}\frac{d{\sigma_{xy}^\tau}}{d{\mu}}
\end{equation}
and
\begin{equation}
\kappa_{xy}^\tau=\frac{\pi^2k_B^2T}{3e^2}\sigma_{xy}^\tau.
\end{equation}

Before discussing the numerical results, we now focus on some analytical results for the irradiated dice lattice ($\alpha=1$) obtained at very low temperature. When the chemical potential lies in the conduction band($+$)/valence band($-$), we find 
\begin{eqnarray}\label{AHC_CB}
\sigma_{xy}^{\pm}=\mp\frac{e^2}{h}\Bigg(1-\frac{\tilde{\Delta}}{\sqrt{\mu^2+\tilde{\Delta}^2}}\Bigg).
\end{eqnarray}
When $\mu$ falls in the band gap, the AHC becomes $\sigma_{xy}^0=\frac{e^2}{h}$. When $\mu$ lies within either conduction band or valence band, we find $\alpha_{xy}^\pm=\pm\frac{\pi k_B e T\tilde{\Delta}}{6 h \mu^2}$, and 
$\alpha_{xy}=0$ otherwise.

\begin{figure}[h!]
\centering
 \includegraphics[width=9.5cm, height=6cm]{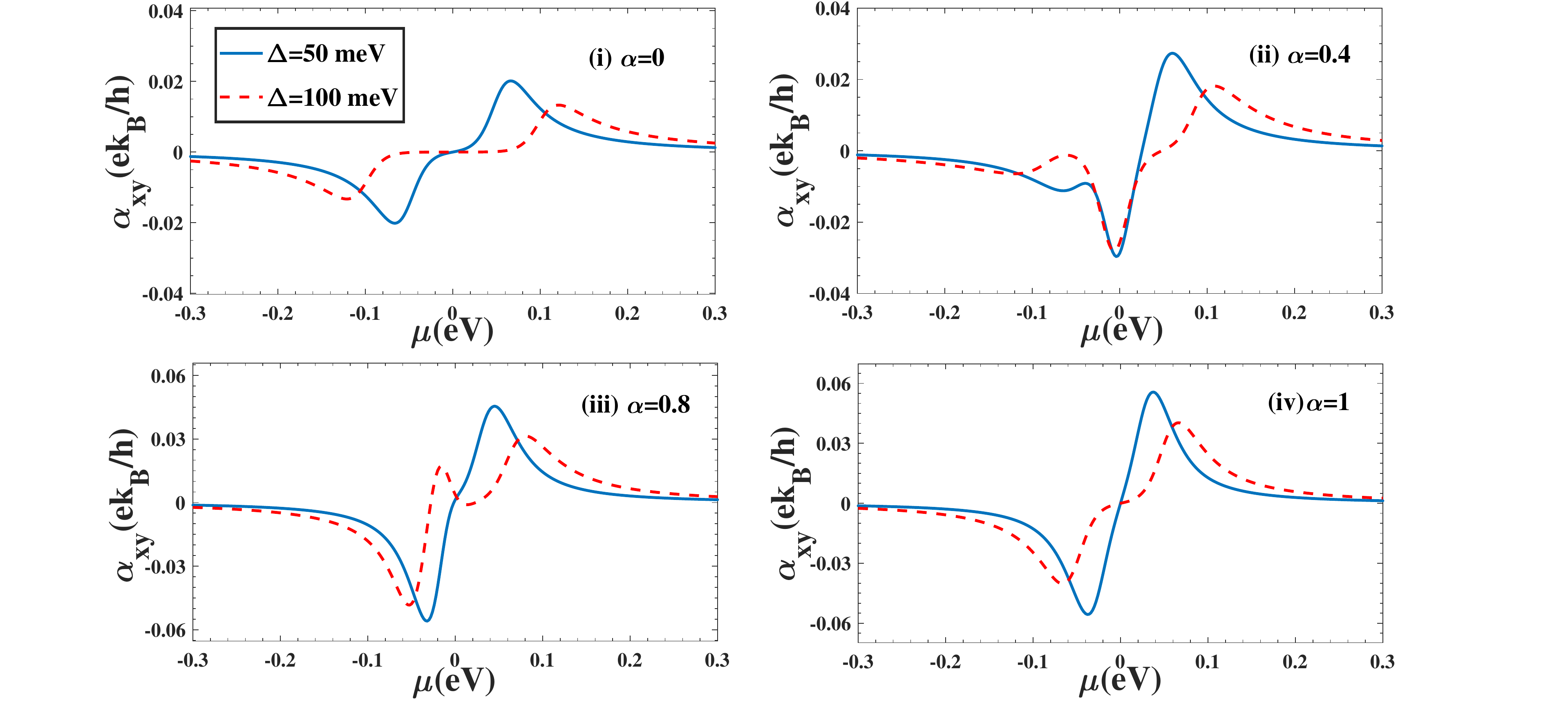}
 \caption{(\textcolor{blue}{Color Online}) The variation of $\alpha_{xy}$ as a function of $\mu$ for various values of $\alpha$ at the $K$-valley. Here, we consider two values of $\Delta$, namely, 50 meV (Solid blue) and 100 meV (Dashed red). We also consider $l=+1$ and $T=100$ K.}
\label{fig:Fig_NernstK}
\end{figure}

The ANC $\alpha_{xy}$ is evaluated numerically from Eq.\,(\ref{ANC}) and its dependence on the chemical potential $\mu$ at 
$T=100$ K is depicted in Fig.\ref{fig:Fig_NernstK} for the $K$- valley. A higher $\Delta$ reduces the magnitude of $\alpha_{xy}$ for all values of $\alpha$. It also causes a shift in the position of the peak towards higher values of $\mu$. It is evident from  
Fig.\,\ref{fig:Fig_NernstK}(i) and Fig.\,\ref{fig:Fig_NernstK} (iv), as 
$\mu$ is varied from the valence band to the conduction band, $\alpha_{xy}$  shows anti-symmetric behavior with a zero value plateau in the band gap for $\alpha=0$ and $\alpha=1$, respectively. However, the plateau corresponding to $\Delta=50$ meV  and $\alpha=1$ is not visible[Fig.\,\ref{fig:Fig_NernstK} (iv)] because the higher thermal energy washes away it in this particular case. These plateaus would be more noticeable at lower temperatures. The width of the plateau is proportional to the photo-induced band gap. The width of the plateau for $\alpha=1$ becomes half of that for $\alpha=0$. This is due to the fact that the photo-induced band gap for the dice lattice is exactly half of that for an irradiated graphene. The vanishing of $\alpha_{xy}$ in the forbidden gap is connected with both the entropy density 
$S_\lambda^\tau(\bm k)$ and the Berry curvature $\Omega(\bm k)$. As depicted in Fig.\,\ref{fig:Fig_BC_K}, $\Omega(\bm k)$ is  mostly concentrated in the band gap at $\bm k=0$ and dying out on either side. On the other hand, at very low temperature, $S_\lambda^\tau(\bm k)$ is sharply peaked at the Fermi surface and vanishes for completely filled and completely empty bands. For $\mu$ slightly above and below the band gap, the intersection of the Fermi surface and the states with non-zero Berry curvature yield a finite contribution to $\alpha_{xy}$ which differs in a sign for $\mu$ below and above the band gap due to the sign change of the Berry curvature. As one approaches the band gap from either side, the Berry curvature starts growing and attains a sharp peak in the band gap near $\bm k=0$, however, the entropy density carries no weight resulting in a vanishing $\alpha_{xy}$ in the band gap.

\begin{figure}[h!]
\centering
 \includegraphics[width=15.5 cm, height=8cm]{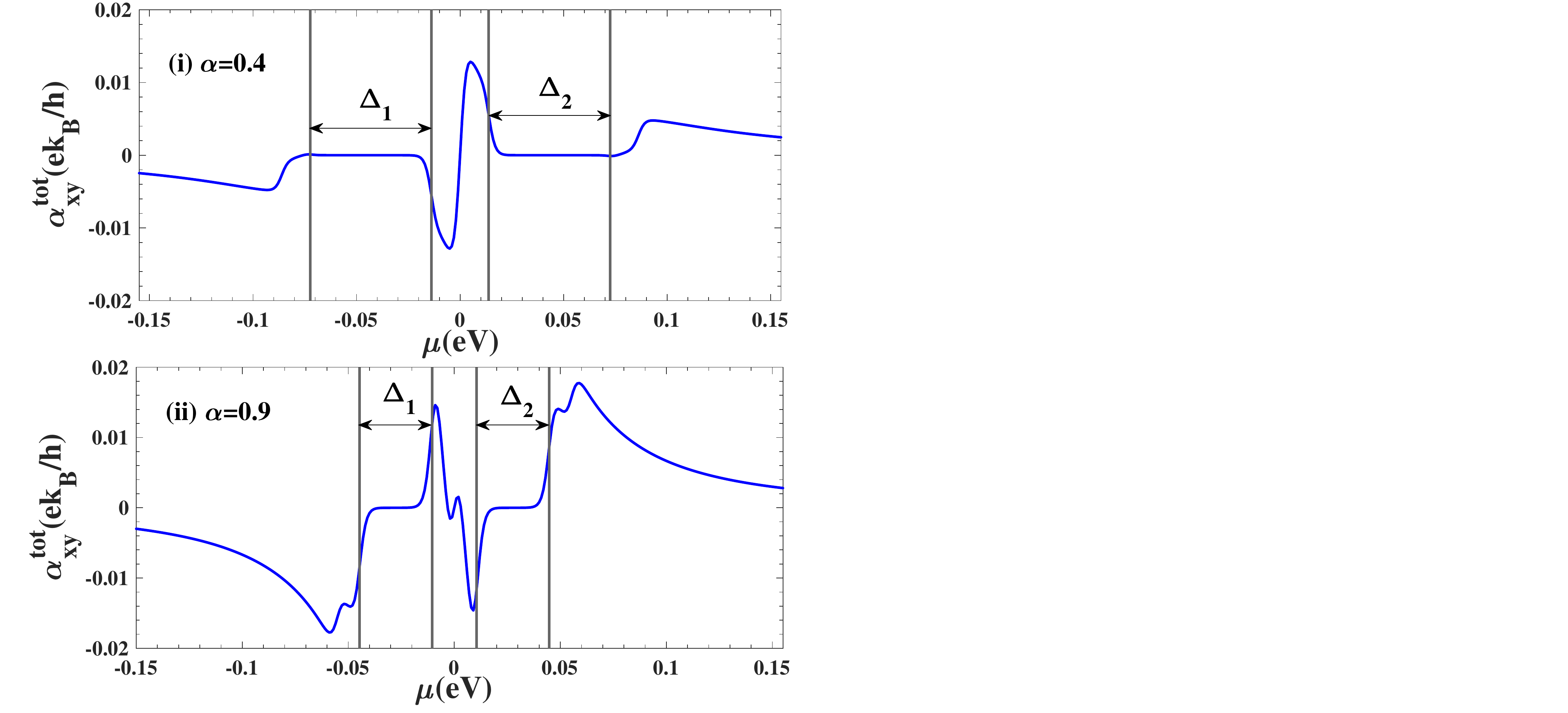}
 \caption{(\textcolor{blue}{Color Online}) Plot of $\alpha_{xy}^{\rm tot}$ versus $\mu$ at $T=20$ K for the $K$-valley. Here, we consider $\Delta=100$ meV and $l=+1$. $\alpha_{xy}^{\rm tot}$ vanishes in the forbidden gaps $\Delta_1$ and $\Delta_2$. For 
$\alpha=0.4$, $\alpha_{xy}^{\rm tot}$ changes sign once from negative to positive when $\mu$ is varied in the region between $\Delta_1$ and $\Delta_2$. However, $\alpha_{xy}^{\rm tot}$ changes its sign several times when $\alpha=0.9$.}
\label{fig:Fig_Nernst15tot}
\end{figure}

The broken particle-hole symmetry corresponding to an intermediate $\alpha(\neq 0,1)$ makes the scenario more interesting. In this case, additional peaks/dips appear in $\alpha_{xy}$ as $\mu$ scans the energy bands [see Fig.\ref{fig:Fig_NernstK}(ii) and Fig.\ref{fig:Fig_NernstK}(iii)]. The plateau(s) in the band gap(s) will be prominent at higher $\Delta$ and lower temperature. 
In Fig.\,\ref{fig:Fig_Nernst15tot}, we show the $\mu$ dependence of total ANC $\alpha_{xy}^{\rm tot}$ i.e sum of individual contributions from both valleys
for (i) $\alpha=0.4$ and (ii) $\alpha=0.9$ considering $ T=20$ K and $\Delta=100$ meV.
In Fig.\,\ref{fig:Fig_Nernst15tot}(i), we notice that $\alpha_{xy}^{\rm tot}$ remains at zero in two distinct ``windows" of $\mu$. As mentioned in the discussion of the orbital magnetization, these ``windows" are basically $\Delta_1$ and $\Delta_2$. Near the edges of each ``window", the plateaus in 
$\alpha_{xy}^{\rm tot}$ are smeared out due to finite temperature. The behavior of $\alpha_{xy}^{\rm tot}$ in the region between 
the ''windows" for $\alpha=0.4$ and $\alpha=0.9$ are completely different. For $\alpha=0.4$, $\alpha_{xy}^{\rm tot}$ changes its sign once from negative to positive. On the other hand, it changes several times when $\alpha=0.9$.

\begin{figure}[h!]
\centering
 \includegraphics[width=10 cm, height=7cm]{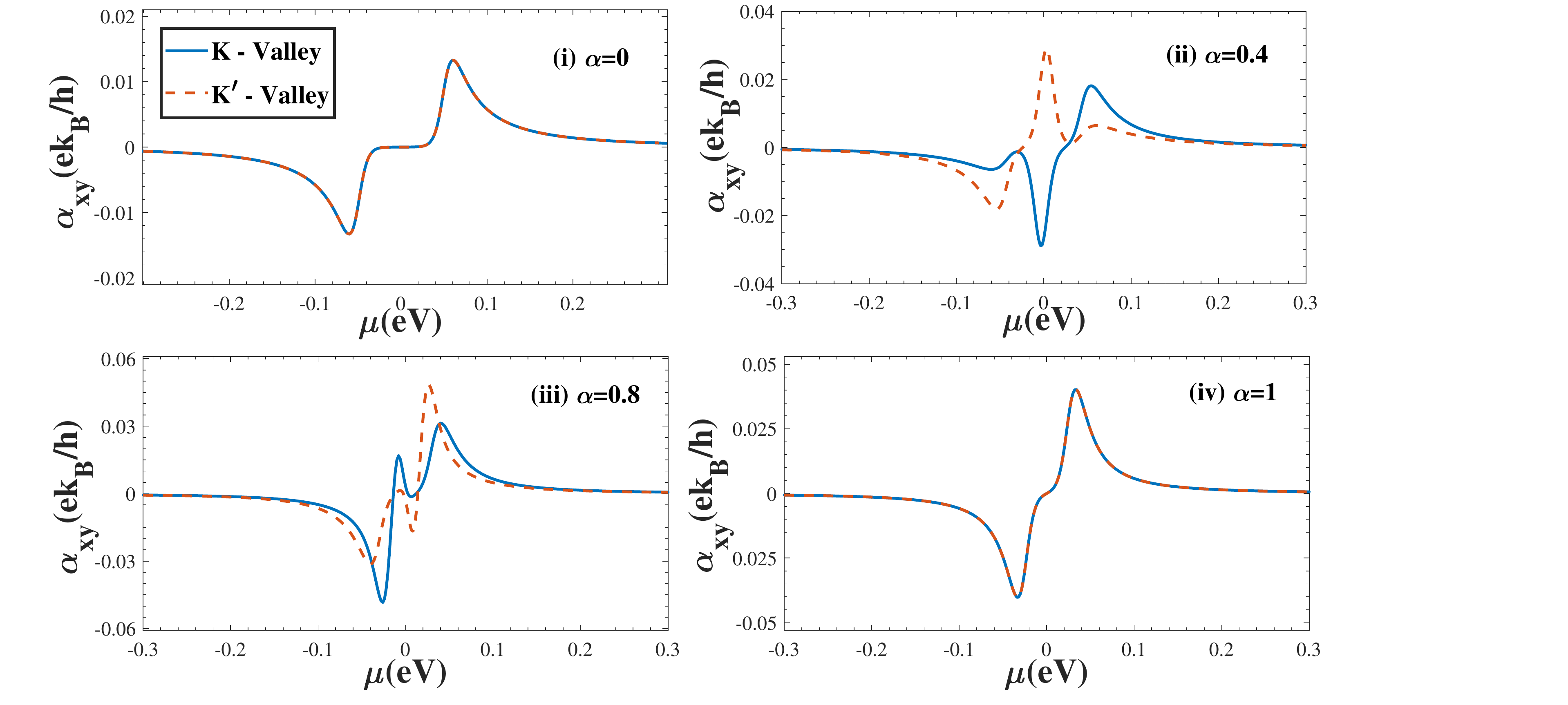}
 \caption{(\textcolor{blue}{Color Online}) Plot of $\alpha_{xy}$ versus $\mu$ at $T=50$ K for both valleys. Here, we consider $\Delta=50$ meV and $l=+1$. $\alpha_{xy}$ exhibits valley contrasting physics at an intermediate $\alpha$ i.e. $0<\alpha<1$ due to broken particle-hole symmetry.}
\label{fig:Fig_Nernst_KKp}
\end{figure}

The $\mu$-dependence of $\alpha_{xy}$ is shown for both valleys in 
Fig.\,\ref{fig:Fig_Nernst_KKp} considering $\Delta=50$ meV, $T=50$ K and 
$l=+1$. As depicted in Fig.\,\ref{fig:Fig_Nernst_KKp}(i) $\&$ (iv), the Nernst coefficient is independent of the valley index $\tau=\pm 1$ for
$\alpha=0$ and $1$.	This can be readily understood from Eq.\,(\ref{ANC}) with the aid of Eq.\,(\ref{Berry_C_Dice}) for $\alpha=1$. This is a direct consequence of the particle-hole symmetry and valley degeneracy. However, the valley-contrasting behavior of $\alpha_{xy}$ is revealed for $0<\alpha<1$ [Fig.\,\ref{fig:Fig_Nernst_KKp}(ii) $\&$ (iii)] as a result of broken particle-hole and valley symmetry.

\begin{figure}[h!]
\centering
\includegraphics[width=9 cm, height=6 cm]{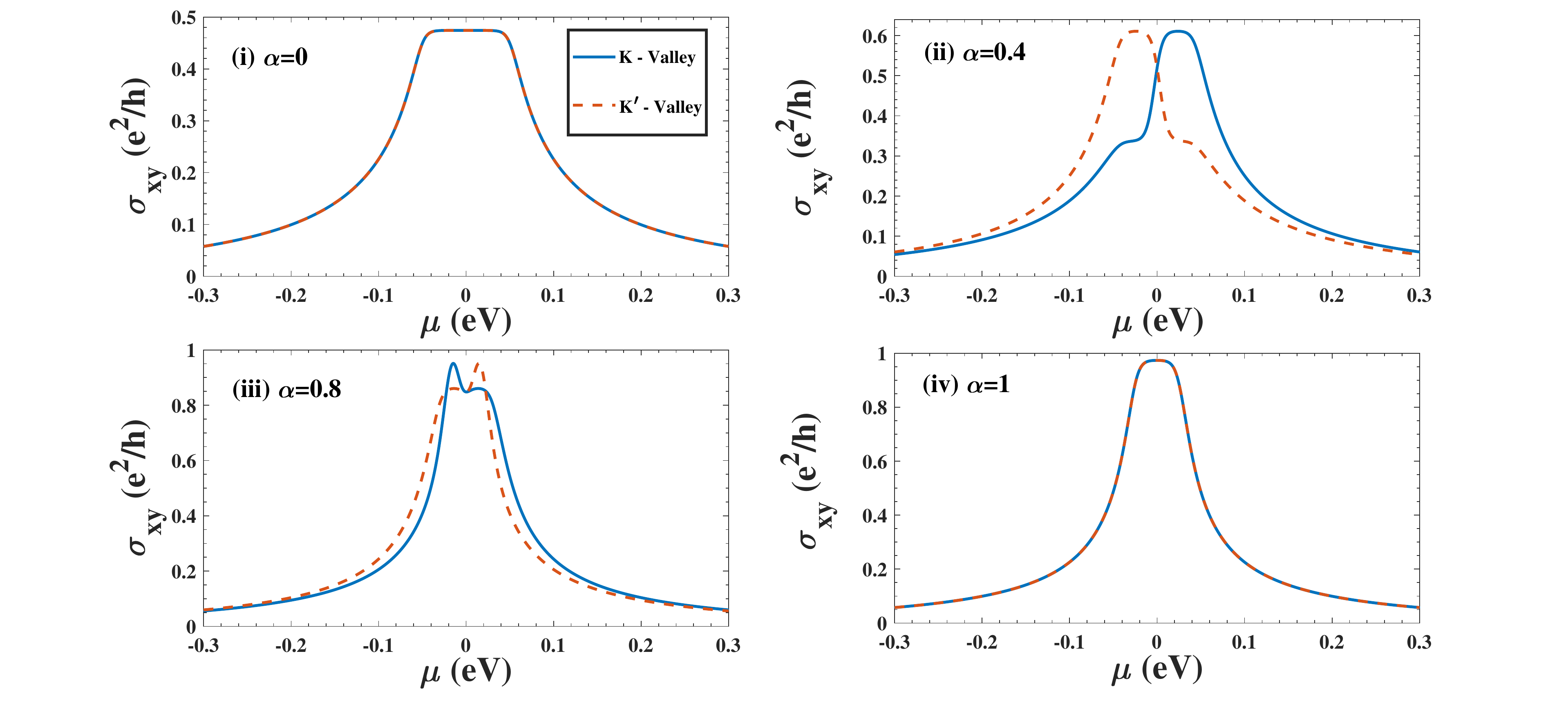}
\caption{(\textcolor{blue}{Color Online}) Plot of $\sigma_{xy}$ versus $\mu$ at $T=50$ K for both valleys. Here, we consider 
$\Delta=50$ meV and $l=+1$. Valley contrasting features in $\sigma_{xy}$ are realized for $0<\alpha<1$.}
\label{fig:Fig_Hall}
\end{figure}

The AHC $\sigma_{xy}$ is calculated numerically from 
Eq.\,(\ref{AHC}) and its variation with the chemical potential at $T=50$ K is shown in Fig.\,\ref{fig:Fig_Hall}. Since the inversion as well as the particle-hole symmetry is preserved for both graphene and the dice lattice, we find that the Hall conductivities for both the valleys coincide when $\alpha=0$ and $\alpha=1$. As $\mu$ is varied in the band gap, in both cases, all the occupied states in the valence band contribute to $\sigma_{xy}$ which results in a plateau of width proportional to $\Delta$. Note that the flat band contributes nothing to $\sigma_{xy}$ because the corresponding Berry curvature vanishes. The height of the plateau for $\alpha=1$ is twice of that corresponding to $\alpha=0$.
For $\alpha\neq0,1$, the system does not possess the inversion and the particle-hole symmetry, resulting in valley contrasting features in the behavior of $\sigma_{xy}$. 
In this case, the "two-plateau" structure is observed as a result of the existence of two band gaps of unequal size i.e. in the quasienergy spectrum at each valley. These features will be more noticeable at lower temperature and higher $\Delta$. The total AHC i.e. sum of the individual contributions from each valley, however, would display interesting features (not shown here explicitly).
As evident from Fig.\,\ref{fig:Fig_Hall}, the total AHC would approach the quantized value $e^2/h\,(2e^2/h)$ approximately in the band gaps $\Delta_1$
and $\Delta_2$ when $\alpha<1/\sqrt{2}\,(\alpha>1/\sqrt{2})$, thus validating a topological phase transition across $\alpha=1/\sqrt{2}$.

\begin{figure}[h!]
\centering
 \includegraphics[width=9.5 cm, height=6cm]{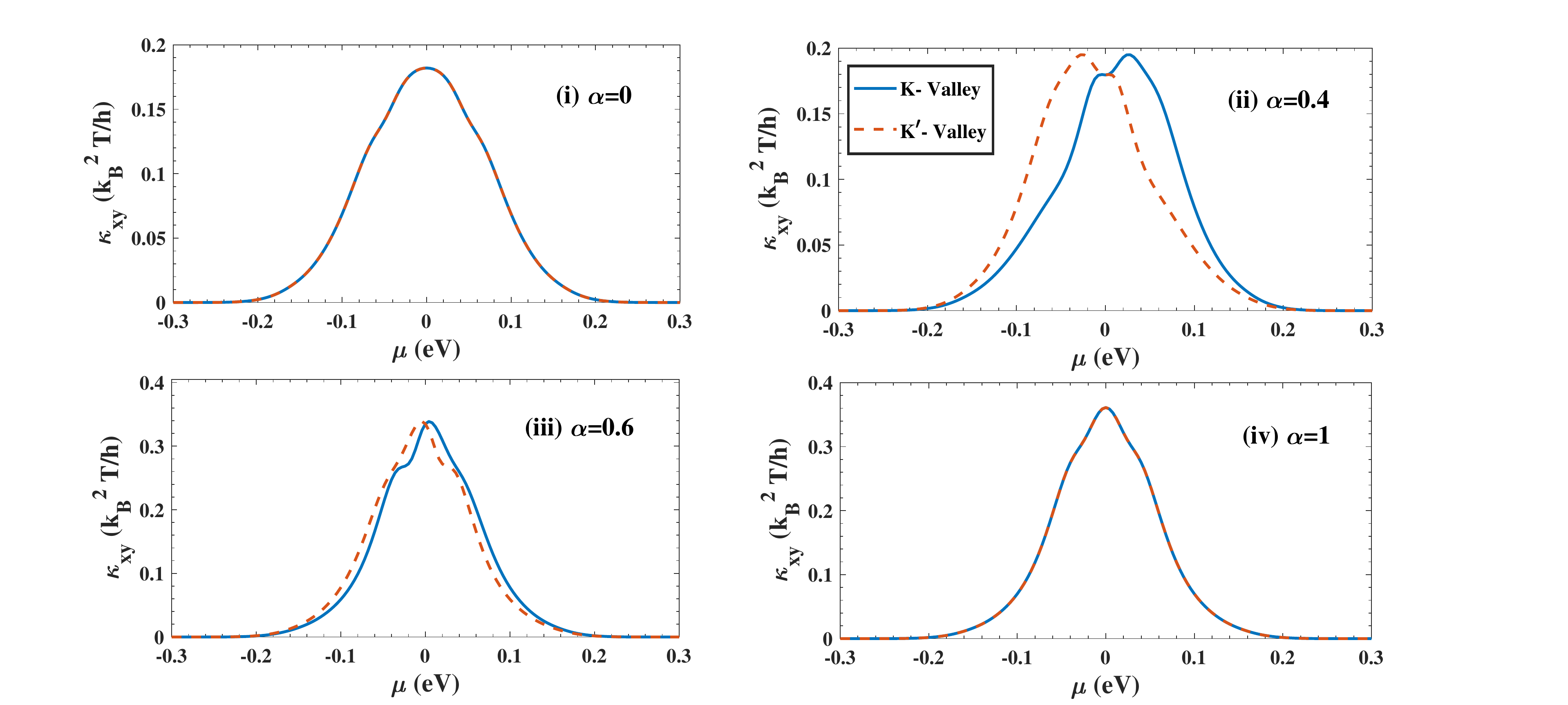}
 \caption{(\textcolor{blue}{Color Online}) Plot of $\kappa_{xy}$ versus $\mu$ at $T=100$ K for both valleys. Here, we consider $\Delta=50$ meV and $l=+1$.}
\label{fig:Fig_Thermal_KKp}
\end{figure}

Using Eq.\,(\ref{THC}), we calculate the THC $\kappa_{xy}$ numerically. Its variation with $\mu$ at both valleys are shown in Fig.\,\ref{fig:Fig_Thermal_KKp} considering $T=100$ K, $\Delta=50$ meV and $l=+1$. For $\alpha=0, 1$, $\kappa_{xy}$ behaves as an even function of $\mu$ unlike $\alpha_{xy}$, owing to the particle-hole symmetry. The THC exhibits similar features as $\sigma_{xy}$ and this similarity would be more prominent at lower temperatures as a validation of the Mott relation. However, 
$\kappa_{xy}$ vanishes away from the band gap regions i.e. deep in the valence band or the conduction band. For $0<\alpha<1$, the valley contrasting features are also available in $\kappa_{xy}$.

\section{Summary}
In summary, we have explored the topological signatures of the irradiated 
$\alpha$-$T_3$ lattice via various Berry phase effects. Explicitly, we calculate the Berry curvature, the orbital magnetic moment, the orbital magnetization and the anomalous thermoelectric coefficients. All these quantities display distinct topological characteristics which can be captured experimentally. The Berry curvature as well as the orbital magnetic moment associated with the flat band display a sign-change across $\alpha=1/\sqrt{2}$. The light induced distortion of the flat band near the Dirac points essentially introduces two well separated $\alpha$-dependent forbidden gaps of equal width. The orbital magnetization exhibits linear dependence on the chemical potential in the forbidden gaps. The slopes of the linear portion in the orbital magnetization for $\alpha>1/\sqrt{2}$ is approximately twice of that for $\alpha<1/\sqrt{2}$. This change in slope is closely related to the transition of the Chern number across 
$\alpha=1/\sqrt{2}$.
The anomalous Nernst coefficient, however, vanishes when the chemical potential is varied in the band gaps. 
The anomalous Hall conductivity attains a plateau whenever the chemical potential falls in the band gap. For $0<\alpha<1$, a ``two-plateau" structure in the Hall conductivity is observed at individual valleys. However, the total anomalous Hall conductivity in the band gaps approaches $e^2/h$ and $2e^2/h$, approximately when  $\alpha<1/\sqrt{2}$ and $\alpha>1/\sqrt{2}$, respectively. 
The thermal Hall conductivity follows the anomalous Hall conductivity in a similar way. For $0<\alpha<1$, the broken particle-hole symmetry introduces the valley contrasting features in the orbital magnetization and the thermoelectric coefficients. These features essentially suggest that the driven $\alpha$-$T_3$ lattice could be used as a potential ingredient in valley caloritronic devices.
We obtain closed analytical expressions of the above mentioned quantities in the case of the irradiated dice lattice $(\alpha=1)$. The analytical results are valley independent owing to the particle-hole as well as the inversion symmetry. The Berry curvature associated with the flat band vanishes whereas the flat band contributes a significant amount to the orbital magnetic moment. Moreover, the contribution of the flat band in the orbital magnetic moment is the sum of individual contributions coming from the conduction and the valence bands. 
\\
\section{Acknowledgement}
T. B. sincerely acknowledges the financial support provided by the University of North Bengal
through University Research Projects to pursue this work.

\end{document}